\documentclass[journal]{IEEEtran}  
\usepackage{amsmath,graphicx,amssymb,amsfonts}
\usepackage{bm}
\usepackage{cite}
\usepackage{algorithm}
\usepackage{algorithmic}
\usepackage{caption}
\usepackage{subcaption}
\usepackage{balance}
\usepackage[dvipsnames]{xcolor}
\usepackage{xcolor}
\usepackage{array}
\usepackage{textcomp}

\begin{document}
	
	\title{Cramér-Rao Bound Optimization for Joint Beamforming and Mode Selection in RDARS-Assisted ISAC Systems}

	\author{Ahmad Reza Hassanshahi, Rouhollah Amiri, Fereidoon Behnia }
	
	\maketitle
	
	\begin{abstract}
		
Integrated Sensing and Communication (ISAC) is a foundation of 6G networks, demanding architectures that simultaneously enhance sensing accuracy and communication reliability. This paper presents a Reconfigurable Distributed Antenna and Reflecting Surface (RDARS) aided ISAC framework, where an RDARS overcomes the limitations of conventional passive Reconfigurable Intelligent Surfaces (RIS) and Distributed Antenna Systems (DAS). By enabling each element to dynamically operate in either \textit{reflection} or \textit{connection} mode, RDARS synergistically harnesses reflection gain, distribution gain, and an additional mode-selection gain. We investigate the joint optimization of transmit beamforming at the base station and dynamic mode selection at the RDARS to minimize the sensing performance metric, namely the Cram\'er-Rao Bound (CRB) for target localization, while guaranteeing a minimum required Signal-to-Interference-plus-Noise Ratio (SINR) for multiple communication users. To solve the resulting non-convex and mixed-integer problem, we develop an efficient iterative algorithm based on the Alternating Optimization (AO) framework, effectively leveraging Majorization-Minimization (MM) and Penalty methods. Comprehensive simulations validate the proposed design, demonstrating that the dynamic RDARS configuration achieves a superior trade-off between the Position Error Bound (PEB) and communication SINR, significantly outperforming benchmark passive RIS and DAS systems.

\end{abstract}

\section{Introduction} \label{sec:1}

\IEEEPARstart{T}{he} relentless advancement towards 6G networks is characterized by a strategic migration to higher frequency bands, such as mmWave and Terahertz. This transition unlocks a paradigm where radio signals transcend their traditional role, facilitating not only high-speed data transmission but also the acquisition of high-resolution information about the surrounding environment \cite{saad2020vision,letaief2019roadmap,tataria20216g,zhang2021perceptive}. This convergence is formalized in Integrated Sensing and Communication (ISAC), a key enabling technology that effectively optimizes spectral and hardware resources by consolidating sensing and communication functionalities \cite{liu2022integrated,meng2025integrated,zhang2025integrated2,liu2020joint,keskin2025fundamental,zhang2022enabling}. Emerging applications, including autonomous navigation, smart cities, and industrial automation, demand not merely connectivity but an elevated capacity for environmental perception and comprehension \cite{liu2022integrated}. ISAC is thus poised to be integral to 6G, providing the critical real-time environmental awareness required for these applications \cite{liu2022integrated}. Furthermore, architectures like the Distributed Antenna Systems (DAS) and its evolutions (e.g., cell-free massive MIMO), which distribute network elements to reduce access distance, can provide a foundational infrastructure conducive to the deployment of sophisticated ISAC frameworks, despite challenges related to power and complexity \cite{ammar2021user,elhoushy2021cell}.

\subsection{Related Works}
The design of ISAC systems fundamentally involves a trade-off between communication performance and sensing accuracy, often quantified by distinct metrics. On the communication side, common objectives are the maximization of the sum rate \cite{zhang2020capacity,ning2023unified} or the guaranteeing of a minimum Signal-to-Interference-plus-Noise Ratio (SINR) for each user \cite{wang2022noma,zhi2022power}. Conversely, sensing performance is frequently evaluated through the maximization of radar Signal-to-Clutter-plus-Noise Ratio \cite{yu2022ris,liu2022radar}, beamforming gain towards targets \cite{liu2018toward,zhang2025low}, or the minimization of the Cramér-Rao Bound (CRB) \cite{liu2021cramer,song2023intelligent}, which provides a lower bound on the variance of unbiased estimators for sensing parameters like target direction or distance. The system architecture plays a pivotal role in managing these trade-offs. While DAS and their evolutions offer a \textit{distribution gain}, they incur high power consumption and hardware costs \cite{ammar2021user,elhoushy2021cell}. Reconfigurable Intelligent Surface (RIS) has emerged as a low-cost alternative to reshape the wireless environment \cite{liu2021reconfigurable,wu2019towards}. Subsequent variants like Simultaneously Transmitting and Reflecting RIS (STAR-RIS) \cite{xu2021star,mu2021simultaneously} and Hybrid Relay-RIS (HR-RIS) \cite{zhang2023active,long2021active} have been developed to overcome inherent limitations such as limited coverage and the multiplicative fading effect. Recent research has extensively explored the integration of these surfaces into ISAC, focusing on joint beamforming design for both fully-passive RIS and active/hybrid variants to balance communication and sensing objectives \cite{liu2021cramer,song2023intelligent,yu2023active,zhu2023joint,zhong2023joint,liu2023snr,chen2023simultaneous,qian2023sensing}. To harness the benefits of both DAS and RIS, the novel Reconfigurable Distributed Antennas and Reflecting Surfaces (RDARS) architecture was proposed in \cite{ma2024reconfigurable}. The RDARS, whose elements operating in either connected or reflected mode, achieves a \textit{hybrid distribution and reflection gain}. After the preliminary work in \cite{ma2024reconfigurable}, RDARS has begun to explore its potential by designing an active/passive beamformer in MIMO communication systems \cite{wang2024joint,zhang2025rate} and investigating the trade-off between rate and SNR in an RDARS-aided ISAC system \cite{zhang2025integrated}.
However, a critical gap remains unaddressed: the integration of RDARS within an ISAC framework optimized for sensing accuracy via the CRB. To the best of our knowledge, no existing work has jointly optimized transmit beamforming and RDARS configuration to minimize the position CRB for target localization under stringent communication QoS constraints. While \cite{zhang2025integrated} studies RDARS-aided ISAC using sensing SINR, our work adopts the position-domain CRB as the localization metric. Recent closely related developments include an RDARS ISAC/localization demonstration, hybrid-RIS mode-selection designs, active-RIS CRB optimization, and tri-mode RDARS-ISAC waveform/mode switching \cite{wang2023demo,lin2025hybrid,zhang2026activecrb,ji2026modeswitch}. None of these works addresses the joint connected/reflected RDARS mode selection and transmit design for position-CRB minimization under multi-user SINR constraints. In the proposed formulation, the RDARS configuration changes both the communication channel and the geometry-dependent Fisher information contributed by the BS/RDARS-target paths. This localization-oriented coupling, rather than the use of standard AO/MM tools by itself, is the main technical distinction from prior RDARS-ISAC designs.

\subsection{Contributions}
\label{subsec:contributions}

This paper jointly optimizes transmit beamforming and dynamic mode selection in an RDARS-aided ISAC system to minimize the CRB while ensuring multi-user SINR. The NP-hard problem, involving variable coupling and unit-modulus/binary constraints, is addressed via an Alternating Optimization (AO) framework combining relaxation, successive approximation/MM, and penalty methods. Simulations confirm its superiority over passive RIS and DAS benchmarks in beampattern, geometry, and Position Error Bound (PEB)-SINR trade-off. The main contributions are:

\begin{itemize}
    \item Tractable SINR and CRB expressions as functions of active beamforming \(\mathbf{F}\), reflection coefficients \(\boldsymbol{\Phi}\), and mode selection \(\mathbf{A}\), enabling a concrete problem formulation.
    
    \item An efficient AO-based iterative algorithm decomposing the problem into sub-problems, with MM/successive approximation handling non-convexities and penalty terms enforcing SINR and mode-selection consistency.
    
    \item Extensive simulations evaluating beampatterns, geometry, and sensing-communication trade-off. The proposed RDARS outperforms passive RIS and DAS, achieving lower PEB under same SINR constraints by exploiting hybrid \textit{distribution} and \textit{reflection} gains.
\end{itemize}

\subsection{Outline}
The remainder of this paper is organized as follows. Section II presents the RDARS-assisted ISAC system model, including communication and radar channels, and defines the SINR and CRB metrics. Section III derives the CRB via the FIM, establishing the theoretical lower bound for localization accuracy. Section IV formulates the joint beamforming and mode selection problem and develops an AO-based solution integrating MM and penalty techniques. Section V provides simulation results, including beampatterns, trade-off analyses, geometric evaluations, and parametric studies, to validate the superiority of the proposed scheme over benchmarks in terms of PEB and SINR. Finally, Section VI concludes the paper.

\textit{Notations:} Scalars, vectors, and matrices are denoted by italic, bold lowercase, and bold uppercase letters, respectively. For a matrix $\mathbf{A}$, $(\cdot)^{T}$, $(\cdot)^{*}$, $(\cdot)^{H}$, and $\operatorname{tr}(\cdot)$ denote transpose, conjugate, conjugate-transpose, and trace; $\mathbf{A}(i,j)$ denotes its $(i,j)$-th entry, and $\|\mathbf{A}\|_{F}$ is the Frobenius norm. $\operatorname{diag}(\mathbf{a})$ forms a diagonal matrix from vector $\mathbf{a}$, $\mathrm{blkdiag}(\cdot)$ forms a block-diagonal matrix, and $\operatorname{vec}(\cdot)$ denotes vectorization. For a complex number, $|\cdot|$ is the modulus and $\|\cdot\|$ the Euclidean norm; $\Re\{\cdot\}$ and $\Im\{\cdot\}$ extract real and imaginary parts. $\mathbb{E}\{\cdot\}$ denotes statistical expectation, and $\mathcal{CN}(\boldsymbol{\mu}, \mathbf{\Sigma})$ represents a CSCG distribution. $\mathbb{R}^{a\times b}$ and $\mathbb{C}^{a\times b}$ denote the sets of $a\times b$ real and complex matrices, respectively. $\mathbf{I}$ and $\mathbf{O}$ are the identity and zero matrices, whose dimensions are clear from the context.

\section{Problem Formulation And System Model}\label{sec:2}

\subsection{Scenario Description}

\begin{figure}
    	\centering
	\centerline{\includegraphics[width=9.5cm]{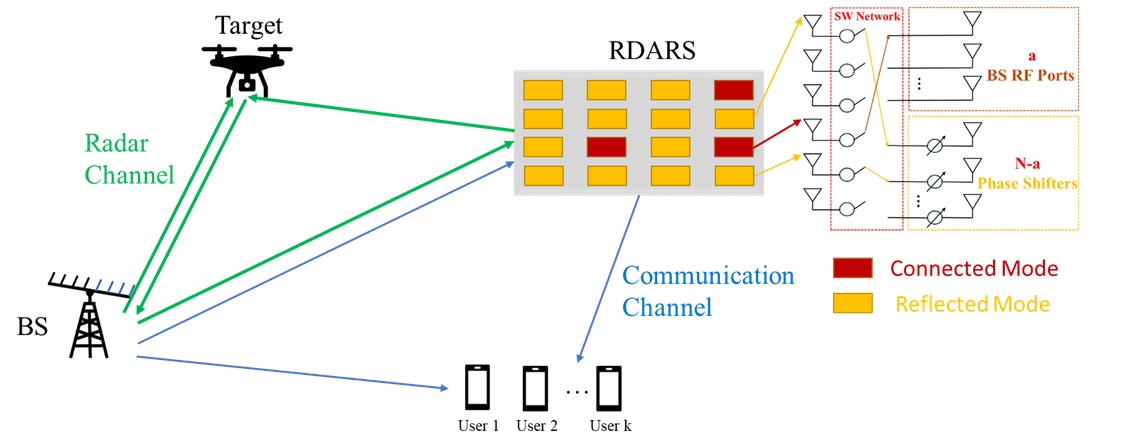}}
\caption[The system model]{\small The system model of an RDARS-aided ISAC system.}
        \label{fig:System-Model}
    \end{figure}

We consider an ISAC system comprising a Base Station (BS), $K$ single-antenna users, a target and an RDARS panel, operating under MIMO-OFDM signaling with \(N_s\) subcarriers and the subcarrier spacing \(\Delta f\). We use centered frequency indices $n\in\mathcal{N}_s\triangleq\{-(N_s-1)/2,\ldots,(N_s-1)/2\}$; for even $N_s$, these are half-integer offsets with respect to the band center. The objective is to accurately sense and locate the target while ensuring user communications remain at a desirable level. The performance is evaluated using the communication SINR and the radar CRB metrics. A schematic of the scenario is shown in Fig. \ref{fig:System-Model}.

The BS, equipped with $M_t=M_r=M$ antennas arranged in a Uniform Linear Array (ULA), concurrently serves $K$ users and detects a point target. The RDARS consists of $N$ elements arranged in a Uniform Planar Array (UPA) configuration, each dynamically switchable between two modes: \emph{reflected mode} (passive phase shifting) and \emph{connected mode} (active transmission and reception). Only $a \ll N$ elements operate in connected mode for hardware efficiency. The binary diagonal matrix $\mathbf{A} = \mathrm{diag}\{a_1, \dots, a_N\}$ indicates the mode of each element ($a_i = 1$ for connected, $a_i = 0$ for reflected). The reflection coefficients are modeled by $\boldsymbol{\Phi} = \mathrm{diag}\{\phi_1, \dots, \phi_N\}$, where $\phi_i=|\phi_i|e^{j\mathrm{arg}(\phi_i)}$ with $\mathrm{arg}({\phi}_{i})\in[0,2\pi)$ and $|\phi_i| = 1$. The overall RDARS reflection effect is captured by $(\mathbf{I} - \mathbf{A})\boldsymbol{\Phi}$. In downlink, the BS and connected RDARS elements act as a distributed antenna system. The transmitted signal is
\begin{equation}
\mathbf{X}[n] = \scalebox{0.9}{$\begin{bmatrix} \mathbf{x}_1[n] \\ \mathbf{x}_2[n] \end{bmatrix} = \begin{bmatrix} \mathbf{F}_1 \\ \mathbf{F}_2 \end{bmatrix} \mathbf{s}[n] \triangleq \mathbf{F}\mathbf{s}[n]$}, \quad \scalebox{0.9}{$\forall n\in\mathcal{N}_s$}.
\label{eq:transmitted_signal}
\end{equation}
where $\mathbf{s}[n] \sim \mathcal{CN}(\mathbf{0}, \mathbf{I}_K)$ contains user data symbols, and $\mathbf{F}_1 \in \mathbb{C}^{M \times K}$ and $\mathbf{F}_2 \in \mathbb{C}^{a \times K}$ are beamforming matrices at the BS and RDARS connected elements, respectively, and we define $\mathbf{R}_{X1}$ and $\mathbf{R}_{X2}$ as follows: 
\begin{equation}
\mathbf{R}_{X1} = \scalebox{0.9}{$\mathbf{F}_1 \mathbf{F}_1^H, \quad \quad 
\mathbf{R}_{X2} = \mathbf{F}_2 \mathbf{F}_2^H$}.
\label{eq:R_x}
\end{equation}

\subsection{Communication Channel}

The communication channels in the considered system are defined as follows. Let $\mathbf{H}_{br} \in \mathbb{C}^{N \times M}$ represent the channel matrix between the BS and RDARS, $\mathbf{h}_{bu,k} \in \mathbb{C}^{M}$ denote the channel vector from the BS to the $k$-th user, and $\mathbf{h}_{ru,k} \in \mathbb{C}^{N}$ signify the channel vector from the RDARS to the $k$-th user. All channel components are characterized by a general Rician fading model. Specifically, the BS-RDARS channel $\mathbf{H}_{br}$ is expressed as
\begin{equation}
\mathbf{H}_{br} = \scalebox{0.9}{$\sqrt{\alpha_{br}}\left( \sqrt{\kappa} \, \mathbf{H}_{br}^{\text{LoS}} + \sqrt{1 - \kappa} \, \mathbf{H}_{br}^{\text{NLoS}} \right)$},
\label{eq:channel_model}
\end{equation}

\noindent where $\alpha_{br}$ is the path loss coefficient, $\kappa$ denotes the Rician factor, $\mathbf{H}_{br}^{\text{NLoS}}$ represents the non-line-of-sight (NLoS) component whose entries are independent and identically distributed (i.i.d.) complex Gaussian random variables, i.e., $\mathbf{H}_{br}^{\text{NLoS}} (i,j) \sim \mathcal{CN}(0,1)$, and $\mathbf{H}_{br}^{\text{LoS}}$ is the deterministic line-of-sight (LoS) component.

%The LoS component is formulated as $\mathbf{H}_{br}^{\text{LoS}} = \mathbf{a}_{\text{r}}(\theta_{br}^{r}, \psi_{br}^{r}) \mathbf{a}_{\text{t}}^{T}(\theta_{br}^{t})$, where $\mathbf{a}_{\text{t}}(\theta_{br}^{t})$ is the transmit array response vector of the BS, and $\mathbf{a}_{\text{r}}(\theta_{br}^{r}, \psi_{br}^{r})$ is the receive array response vector of the RDARS. The angles $\theta_{br}^{t}$, $\theta_{br}^{r}$, and $\psi_{br}^{r}$ correspond to the azimuth angle of departure (AoD) at the BS, the azimuth angle of arrival (AoA) at the RDARS, and the elevation AoA at the RDARS, respectively. The array response vectors for an $M$-element Uniform Linear Array (ULA) and an $N_1 \times N_2$ Uniform Planar Array (UPA) (with $N = N_1 N_2$) are respectively given by:

%\begin{align}
%\mathbf{a}_{\text{ULA}}(\theta) &= \left[ 1, e^{-j\frac{2\pi}{\lambda}d \sin(\theta)}, \ldots, e^{-j\frac{2\pi}{\lambda}d(M-1)\sin(\theta)} \right]^{T}, \label{eq:ula_response} \\
%\mathbf{a}_{\text{UPA}}(\theta, \psi) &= \left[ 1, e^{-j\frac{2\pi}{\lambda}d( n_x \cos(\theta)\sin(\psi) + n_y \sin(\psi) )}, \ldots, \right. \nonumber \\
%&\qquad \left. e^{-j\frac{2\pi}{\lambda}d( (N_1-1)\cos(\theta)\sin(\psi) + (N_2-1)\sin(\psi) )} \right]^{T}, \label{eq:upa_response}
%\end{align}

%\noindent where $\theta$ and $\psi$ represent the azimuth and elevation angles, respectively, $\lambda$ is the carrier wavelength, and $d$ is the inter-element spacing, typically set to $d = \lambda/2$.

Considering the described system model, the received signal at the $k$-th user can be formulated as:
\begin{align}
\mathbf{Y}_{u,k}[n] &= \scalebox{0.8}{$\displaystyle \left( \mathbf{h}_{bu,k}^{T} + \mathbf{h}_{ru,k}^{T} (\mathbf{I} - \mathbf{A}) \boldsymbol{\Phi} \mathbf{H}_{br} \right) \mathbf{x}_{1}[n] + \mathbf{h}_{ru,k}^{T} \mathbf{A}_{a} \mathbf{x}_{2}[n] + \mathbf{Z}_{C,k}[n]$} \nonumber \\
&= \scalebox{0.95}{$\mathbf{h}_{k}^{T} \mathbf{F} \mathbf{s}[n] + \mathbf{Z}_{C,k}[n]$}, \label{eq:received_signal}
\end{align}

\noindent where $\mathbf{h}_{k} = \left[ \mathbf{h}_{bu,k}^{T} + \mathbf{h}_{ru,k}^{T} (\mathbf{I} - \mathbf{A}) \boldsymbol{\Phi} \mathbf{H}_{br}, \quad \mathbf{h}_{ru,k}^{T} \mathbf{A}_{a} \right]^{T}$ represents the composite communication channel from the BS and the RDARS's connected elements to the $k$-th user, and $\mathbf{Z}_{C,k}[n] \sim \mathcal{CN}(\mathbf{0}, \mathbf{\sigma}^2_k)$ is the additive white Gaussian noise (AWGN). It is important to highlight the introduction of an additional selection matrix $\mathbf{A}_{a} \in \mathbb{R}^{N \times a}$ in \eqref{eq:received_signal}, which is employed to handle the dimensionality mismatch between the signal vectors $\mathbf{x}_1[n]$ and $\mathbf{x}_2[n]$. Specifically, $\mathbf{A}_{a}$ is constructed by extracting the $a$ columns from the master selection matrix $\mathbf{A}$ that correspond to the connected-mode elements. This relationship can be compactly expressed as $\mathbf{A} = \mathbf{A}_a \mathbf{A}_a^{\mathrm{T}}$.

Based on the received communication signal model \eqref{eq:received_signal}, the SINR for the $k$-th user can be formulated as:
\begin{equation}
\textbf{SINR}_{u,k} = \frac{ \left| \mathbf{h}_{k}^{\mathrm{T}} \mathbf{f}_k \right|^2 }{ \sum_{i \neq k}^{K} \left| \mathbf{h}_{k}^{\mathrm{T}} \mathbf{f}_i \right|^2 + \sigma_{k}^2 },
\label{eq:sinr_formula}
\end{equation}

\noindent where $\mathbf{f}_i$ denotes the $i$-th column of the composite beamforming matrix $\mathbf{F}$, for $i = 1, \ldots, K$, representing the beamforming vector dedicated to user $i$.

\subsection{Radar Channel}
\label{subsec:radar_channel}

This section analyzes the received radar signal. As shown in Fig.~\ref{fig:System-Model}, the radar channel comprises five distinct propagation paths:

\begin{enumerate}
    \item \textbf{BS $\to$ Target $\to$ BS Path:} The signal travels directly from the BS to the target and back to the BS.
    
    \item \textbf{BS $\to$ Target $\to$ RDARS Path:} The signal propagates from the BS to the target and is then reflected back to the RDARS's connected elements.
    
    \item \textbf{BS $\to$ Target $\to$ RDARS $\to$ BS Path:} The signal travels from the BS to the target and is then reflected to the RDARS's reflection elements and is finally reflected back to the BS.
    
    \item \textbf{RDARS $\to$ Target $\to$ RDARS Path:} The signal is transmitted from the RDARS's connected elements to the target and returns to the RDARS.
    
    \item \textbf{RDARS $\to$ Target $\to$ BS Path:} The signal transmitted from the connected elements of RDARS to the target and is then reflected back to the BS.
\end{enumerate}

The BS, located at $\mathbf{q} \in \mathbb{R}^{2}$, serves $K$ single-antenna users at $\mathbf{u}\in \mathbb{R}^{2}$ while sensing a point target at $\mathbf{p}\in \mathbb{R}^{2}$, aided by an RDARS at $\mathbf{r}\in \mathbb{R}^{2}$. The total received radar signal for subcarrier $n$ is given by:
\begin{equation}
\mathbf{Y}_R[n] = 
\begin{bmatrix}
\alpha_{1}e^{-j2\pi n \Delta f\tau_{1}}{\mathbf{G}}_{1}\mathbf{x}_1[n]
\\
+\alpha_{3}e^{-j2\pi n \Delta f\tau_{3}}{\mathbf{G}}_{3}\mathbf{x}_1[n] 
\\
+\alpha_{5}e^{-j2\pi n \Delta f\tau_{5}}{\mathbf{G}}_{5}\mathbf{x}_2[n] \\
\alpha_{2}e^{-j2\pi n \Delta f\tau_{2}}{\mathbf{G}}_{2}\mathbf{x}_1[n]
\\
+\alpha_{4}e^{-j2\pi n \Delta f \tau_{4}}{\mathbf{G}}_{4}\mathbf{x}_2[n] 
\end{bmatrix} +\mathbf{Z}_R[n],
\label{eq:radar_signal} \\
\end{equation}
where the path delays are:
\begin{align}
\tau_1 &= \scalebox{0.9}{$\frac{2}{c} \| \mathbf{q} - \mathbf{p} \|_2$}, \label{eq:tau1} \\
\tau_2 &= \tau_5 = \scalebox{0.9}{$\frac{1}{c} \left[ \| \mathbf{r} - \mathbf{p} \|_2 + \| \mathbf{q} - \mathbf{p} \|_2 \right]$}, \label{eq:tau2} \\
\tau_3 &= \scalebox{0.9}{$\frac{1}{c} \left[ \| \mathbf{q} - \mathbf{r} \|_2 + \| \mathbf{r} - \mathbf{p} \|_2 + \| \mathbf{p} - \mathbf{q} \|_2 \right]$}, \label{eq:tau3} \\
\tau_4 &= \scalebox{0.9}{$\frac{2}{c} \| \mathbf{r} - \mathbf{p} \|_2$}. \label{eq:tau4} 
\end{align}
The term \(\Delta f\) denotes the subcarrier spacing and $\mathbf{Z}_{R}[n] \sim \mathcal{CN}(\mathbf{0},\sigma^2\mathbf I)$ represents AWGN. The parameters \(\alpha_k\) denote the complex path gains. Here, \(\theta_1\) is the BS-target angle, \(\theta_2\) is the RDARS-target direction in a global frame, \(\theta_3\) is the BS-RDARS angle in the BS frame, \(\theta_4\) is the RDARS-BS angle in the RDARS frame, and \(\theta_5\) is the same RDARS-target direction expressed in the RDARS local frame. The array orientations are assumed known, so $\theta_5=\theta_2+\psi_0$ for a known frame offset $\psi_0$, and therefore $\partial\theta_5/\partial\theta_2=1$.
The beamforming gain matrices $\mathbf{G}_k$ are defined as follows: $\mathbf{G}_1 = \mathbf{a}(\theta_1) \mathbf{a}^H(\theta_1)$, $\mathbf{G}_2 = \mathbf{A}_a^T \mathbf{b}(\theta_2) \mathbf{a}^H(\theta_1)$, $\mathbf{G}_3 = g \, \mathbf{a}(\theta_1) \mathbf{a}^H(\theta_3)$, $\mathbf{G}_4 = \mathbf{A}_a^T \mathbf{b}(\theta_2) \mathbf{b}^H(\theta_2) \mathbf{A}_a$, and $\mathbf{G}_5 = \mathbf{a}(\theta_1) \mathbf{b}^H(\theta_2) \mathbf{A}_a$.
Where $\mathbf{a}(\cdot)$ denotes the BS array response and $\mathbf{b}_{\text{eff}}(\cdot) = \mathbf{A}_a^T \mathbf{b}_{\text{full}}(\cdot)$ represents the effective array response of the connected RDARS elements. The gain factor \(g\) for path 3 is computed as follows (where \(\mathbf{c}(\cdot)\) corresponds to the RDARS reflected elements array response):
\begin{equation}
g = \mathbf{c}^H(\theta_4) \left[ (\mathbf{I} - \mathbf{A}) \boldsymbol{\Phi} \right] \mathbf{c}(\theta_5).
\label{eq:gain_g}
\end{equation}

\subsection{Performance Metrics}

To evaluate and compare different systems, appropriate performance metrics must be defined. For ISAC systems, these typically fall into two categories: communication metrics (e.g., data rate or SINR) and radar metrics (e.g., target localization accuracy). In this work, we adopt the SINR as the communication metric and the CRB as the radar performance metric. The CRB provides a fundamental lower bound on the covariance matrix of any unbiased estimator \cite{kay1993fundamentals}.
Based on the received signal model in \eqref{eq:radar_signal}, the complete parameter vector \(\boldsymbol{\gamma} \in \mathbb{R}^{12}\) includes both the target position and the path loss coefficients is given by $\boldsymbol{\gamma} = [\mathbf{p}^{T}, \boldsymbol{\alpha}^{T}]^{T}$,
where $\boldsymbol{\alpha} = [\boldsymbol{\alpha}_1^{T}, \boldsymbol{\alpha}_2^{T}, \boldsymbol{\alpha}_3^{T}, \boldsymbol{\alpha}_4^{T}, \boldsymbol{\alpha}_5^{T}]^{T}$ and each sub-vector $\boldsymbol{\alpha}_k$ contains the real and imaginary parts of the path loss coefficient for the $k$-th path, i.e., $\boldsymbol{\alpha}_k = [\Re\{\alpha_k\}, \Im\{\alpha_k\}]^{T}$.

Let $\hat{\boldsymbol{\gamma}}$ be an estimate of $\boldsymbol{\gamma}$ obtained from the observations $\mathbf{Y}_{R}[n]$'s. The Mean Squared Error (MSE) matrix satisfies the inequality \cite{kay1993fundamentals}:
\begin{equation}
\scalebox{0.9}{$\mathbb{E}\left\{ (\hat{\boldsymbol{\gamma}} - \boldsymbol{\gamma})(\hat{\boldsymbol{\gamma}} - \boldsymbol{\gamma})^{T} \right\} \geq \mathbf{J}_{\boldsymbol{\gamma}}^{-1}$},
\label{eq:crb_inequality}
\end{equation}
where \(\mathbf{J}_{\bm{\gamma}}\) is the FIM for \(\boldsymbol{\gamma}\). In particular, for the target position estimate \(\hat{\mathbf{p}}\), (\ref{eq:crb_inequality}) yields a lower bound on the localization MSE:
\begin{equation}
\scalebox{0.9}{$\mathbb{E}\left\{ (\hat{\mathbf{p}} - \mathbf{p})(\hat{\mathbf{p}} - \mathbf{p})^{T} \right\} \geq \textrm{CRB}(\mathbf{p})$}.
\label{eq:position_crb}
\end{equation}

\section{Calculation of the CRB of Target Position}

This section analyzes the fundamental limits of target localization by deriving the corresponding FIM.
Following the received signal model in \eqref{eq:radar_signal}, the complete parameter vector $\boldsymbol{\gamma} \in \mathbb{R}^{12}$ is defined to encapsulate both the target position and the path loss coefficients:
\begin{equation}
\boldsymbol{\gamma} = [\mathbf{p}^{T}, \boldsymbol{\alpha}^{T}]^{T},
\label{eq:parameter_vector}
\end{equation}
where $\boldsymbol{\alpha} = [\boldsymbol{\alpha}_1^{T}, \boldsymbol{\alpha}_2^{T}, \boldsymbol{\alpha}_3^{T}, \boldsymbol{\alpha}_4^{T}, \boldsymbol{\alpha}_5^{T}]^{T}$. Each sub-vector $\boldsymbol{\alpha}_k$ comprises the real and imaginary parts of the path loss coefficient for the $k$-th path, i.e., $\boldsymbol{\alpha}_k = [\Re\{\alpha_k\}, \Im\{\alpha_k\}]^{T}$.

To simplify the analysis, we introduce a reparameterization from $\boldsymbol{\gamma}$ to an intermediate vector $\boldsymbol{\eta} = [\boldsymbol{\tau}^{T}, \boldsymbol{\theta}^{T}, \boldsymbol{\alpha}^{T}]^{T} \in \mathbb{R}^{17}$, where $\boldsymbol{\tau} = [\tau_1, \tau_2, \tau_3, \tau_4, \tau_5]^{T}$ contains the time delays and $\boldsymbol{\theta} = [\theta_1, \theta_2]^{T}$. For the compact closed-form expressions below, we adopt a resolvable-path approximation: distinct paths are separable over the OFDM/array observation aperture, the centered subcarrier indexing above is used, and the relevant delay/gain and angle/gain score inner products vanish. Under this stated approximation,
\begin{equation}
\mathbf{J}_{\boldsymbol{\eta}} = \operatorname{blkdiag}[\mathbf{J}_\tau, \mathbf{J}_\theta, \mathbf{J}_\alpha].
\end{equation}
If these orthogonality conditions are not satisfied, the full FIM must be retained and the position EFIM is obtained directly from the Schur complement in \eqref{eq:schur_complement}. Under the adopted approximation, the required diagonal terms are:
\begin{align}
J_{\tau_{i},\tau_{i}} &= \scalebox{0.81}{$\frac{2}{\sigma^{2}}|\alpha_{i}|^{2}(2\pi \Delta f)^{2}
\mathrm{tr}\{\mathbf{G}_{i}\mathbf{R}_{x1}\mathbf{G}_{i}^{H}\}  \sum_{n=-(N_s-1)/2}^{(N_s-1)/2} n^{2}$}, \quad \scalebox{0.81}{$i=1,2,3$}
\label{eq:J-tau1} \\
J_{\tau_{j},\tau_{j}} &= \scalebox{0.81}{$\frac{2}{\sigma^{2}}|\alpha_{j}|^{2}(2\pi \Delta f)^{2}
\mathrm{tr}\{\mathbf{G}_{j}\mathbf{R}_{x2}\mathbf{G}_{j}^{H}\} \sum_{n=-(N_s-1)/2}^{(N_s-1)/2} n^{2}$}, \quad \scalebox{0.81}{$j=4,5$}
\label{eq:J-tau2} \\
J_{\theta_{1},\theta_{1}} &= \scalebox{0.7}{$\frac{2{N}_{s}}{\sigma^{2}} \Bigg( \sum_{k \in \{1,2,3\}} |\alpha_{k}|^{2} \, \mathrm{tr}\{\dot{\mathbf{G}}_{k,\theta_{1}} \mathbf{R}_{x1} \dot{\mathbf{G}}_{k,\theta_{1}}^{H}\} + |\alpha_{5}|^{2} \, \mathrm{tr}\{\dot{\mathbf{G}}_{5,\theta_{1}} \mathbf{R}_{x2} \dot{\mathbf{G}}_{5,\theta_{1}}^{H}\} \Bigg)$},
\label{eq:J_theta_split_simple} \\
J_{\theta_{2},\theta_{2}} &= \scalebox{0.74}{$\frac{2N_s}{\sigma^{2}}\Big( |\alpha_2|^2\operatorname{tr}\{\dot{\mathbf G}_{2,\theta_2}\mathbf R_{x1}\dot{\mathbf G}_{2,\theta_2}^H\}+|\alpha_3|^2\operatorname{tr}\{\dot{\mathbf G}_{3,\theta_2}\mathbf R_{x1}\dot{\mathbf G}_{3,\theta_2}^H\}$} \nonumber\\
&\quad \scalebox{0.74}{$+|\alpha_4|^2\operatorname{tr}\{\dot{\mathbf G}_{4,\theta_2}\mathbf R_{x2}\dot{\mathbf G}_{4,\theta_2}^H\}+|\alpha_5|^2\operatorname{tr}\{\dot{\mathbf G}_{5,\theta_2}\mathbf R_{x2}\dot{\mathbf G}_{5,\theta_2}^H\}\Big)$},
\label{eq:J-theta2}\\
J_{\alpha_{i},\alpha_{i}} &= \scalebox{0.78}{$\frac{2}{\sigma^{2}}\mathrm{tr}\{\mathbf{G}_{i}\mathbf{R}_{x1}\mathbf{G}_{i}^{H}\}{N}_{s} \mathbf{I}_{2}$}, \quad \scalebox{0.81}{$i=1,2,3$}
\label{eq:J-alpha1} \\
J_{\alpha_{j},\alpha_{j}} &= \scalebox{0.78}{$\frac{2}{\sigma^{2}}\mathrm{tr}\{\mathbf{G}_{j}\mathbf{R}_{x2}\mathbf{G}_{j}^{H}\}{N}_{s} \mathbf{I}_{2}$},\quad \scalebox{0.81}{$j=4,5$}
\label{eq:J-alpha2}
\end{align}
where $\mathbf{G}_i$ are the beamforming gain matrices, and $\dot{\mathbf{G}}_{i,\theta_k}$ denotes the derivative of $\mathbf{G}_i$ with respect to $\theta_k$. In particular, $\dot{\mathbf G}_{3,\theta_2}$ includes the dependence of $g$ on $\theta_5=\theta_2+\psi_0$. Under the same resolvable-path/orthogonality approximation, $J_{\theta_{1},\theta_{2}} = 0$.
To obtain the position EFIM, we must eliminate the nuisance parameters. The full FIM for $\boldsymbol{\gamma}$ is partitioned as:
\begin{equation}
\mathbf{J}_{\boldsymbol{\gamma}} = 
\begin{bmatrix}
\mathbf{J}_{pp} & \mathbf{J}_{p\alpha} \\
\mathbf{J}_{\alpha p} & \mathbf{J}_{\alpha\alpha}
\end{bmatrix},
\end{equation}
where $\mathbf{J}_{pp}$ corresponds to the position parameters, $\mathbf{J}_{\alpha\alpha}$ corresponds to the path gains, and $\mathbf{J}_{p\alpha} = \mathbf{J}_{\alpha p}^T$ represents the coupling between them. The position EFIM is then obtained via the Schur complement:
\begin{equation}
\mathbf{J}_e = \scalebox{0.9}{$ \mathbf{J}_{pp} - \mathbf{J}_{p\alpha} \mathbf{J}_{\alpha\alpha}^{-1} \mathbf{J}_{\alpha p}$}.
\label{eq:schur_complement}
\end{equation}

Under the resolvable-path and score-orthogonality approximation stated above, the position--gain information block vanishes, i.e., $\mathbf{J}_{p\alpha} = \mathbf{O}$, and the Schur complement reduces to:
\begin{equation}
\mathbf{J}_e = \scalebox{0.9}{$\mathbf{J}_{pp} = \boldsymbol{\Xi}^{T} \mathbf{J}_{m} \boldsymbol{\Xi}$},
\end{equation}
where $\mathbf{J}_{\boldsymbol{m}} = \operatorname{blkdiag}[\mathbf{J}_\tau, \mathbf{J}_\theta]$ and the components $\boldsymbol{\Xi}$ and $\mathbf{T}$ (with $\mathbf{T}$ being the Jacobian transformation matrix) are defined in \eqref{eq:T}. The position-related submatrix denoted as effective FIM is given by:
\begin{equation}
\mathbf{J}_e = \scalebox{0.9}{$\boldsymbol{\Xi}^{T} \mathbf{J}_{m} \boldsymbol{\Xi}$} = 
\scalebox{0.9}{$\begin{bmatrix}
J_{1,1} & J_{1,2} \\
J_{1,2} & J_{2,2}
\end{bmatrix}$}.
\label{eq:J_e_matrix}
\end{equation}
Identifiability of the target position requires $\mathbf{J}_e = \mathbf{J}_{pp} - \mathbf{J}_{p\alpha} \mathbf{J}_{\alpha\alpha}^{-1} \mathbf{J}_{\alpha p} \succ 0$. Degeneracy occurs whenever the available delay/angle score vectors fail to span both position dimensions, for example when distinct paths become locally indistinguishable or the transmit design places a sensing null toward the target. Collinearity can degrade conditioning for some geometries, but is not asserted here to be singular in every case. The CRB is well-defined for the simulated configurations for which $\mathbf J_e$ is positive definite (Fig.~\ref{fig:geometry}). Clutter can be modeled as colored noise $\mathbf{R}_n = \sigma^2 \mathbf{I} + \mathbf{R}_c$, extending the framework via $\sigma^2 \mathbf{I} \to \mathbf{R}_c + \sigma^2 \mathbf{I}$. Moreover, dynamic RCS can be addressed via chance constraints, as demonstrated in recent ISAC literature \cite{khalili2024advanced, khalili2026movable}, which enhance robustness to RCS fluctuations. These extensions are identified as future work. Finally, the CRB of target position is given by:
\begin{equation}
\begin{split}
\text{CRB} = \operatorname{tr}(\mathbf{J}_{e}^{-1}) = \frac{J_{1,1} + J_{2,2}}{J_{1,1} J_{2,2} - J_{1,2}^2}
= \frac{\mathbf{r}^H \mathbf{m} + \mathbf{m}^H \mathbf{r}}{\mathbf{r}^H \tilde{\mathbf{B}} \mathbf{r}},
\end{split}
\label{CRB-exp}
\end{equation}
where we have defined the unified vector $\mathbf{r} = [\mathbf{r}_1^T, \mathbf{r}_2^T]^T \in \mathbb{C}^{M^2+a^2}$ with $\mathbf{r}_1 \triangleq \text{vec}(\mathbf{R}_{X1})$ and $\mathbf{r}_2 \triangleq \text{vec}(\mathbf{R}_{X2})$. The parameters $\mathbf{m}$ and $\tilde{\mathbf{B}}$ are defined in \eqref{eq:ND} and \eqref{eq:CRB-end}. The details of the CRB calculation, are provided in Appendix A.

\section{Joint Beamforming and Mode Selection} \label{sec:3}

\subsection{Problem Formulation}

This paper aims to enhance target positioning accuracy while guaranteeing multi-user communication performance. We formulate an optimization problem that minimizes the CRB under strict SINR constraints, jointly optimizing the composite transmit beamforming matrix $\mathbf{F}$, the RDARS reflection matrix $\boldsymbol{\Phi}$, and the selection matrices $\mathbf{A}$ and $\mathbf{A}_{a}$. Using the SINR and CRB analyses in \eqref{eq:sinr_formula} and \eqref{CRB-exp}, the detailed optimization problem is formulated as:
\begin{subequations}
\begin{align}
    \text{(P1)} \quad & \min_{\mathbf{F},\boldsymbol{\Phi},\mathbf{r},\mathbf{A},\mathbf{A}_{a}} \quad  \scalebox{0.9}{$\frac{\mathbf{r}^H \mathbf{m} + \mathbf{m}^H \mathbf{r}}{\mathbf{r}^H \tilde{\mathbf{B}} \mathbf{r}}$} \nonumber\\
    \text{s.t.} \quad
    & \mathbf{r}_1 = \scalebox{0.9}{$\operatorname{vec}\left(\sum_{k=1}^K \mathbf{F}_{1,k}\right)$}, \label{eq:r1-vec}\\
    & \mathbf{r}_2 = \scalebox{0.9}{$\operatorname{vec}\left(\sum_{k=1}^K \mathbf{F}_{2,k}\right)$}, \label{eq:r2-vec}\\
    & \scalebox{0.9}{$\left| \mathbf{h}_{k}^T \begin{bmatrix} \mathbf{f}_{1,k} \\ \mathbf{f}_{2,k} \end{bmatrix} \right|^2 - \overline{\gamma}_{k} \sum_{i \neq k}^K \left| \mathbf{h}_{k}^T \begin{bmatrix} \mathbf{f}_{1,i} \\ \mathbf{f}_{2,i} \end{bmatrix} \right|^2 \geq \overline{\gamma}_{k} \sigma_{k}^2$}, \label{eq:sinr} \\
    & \scalebox{0.9}{$\|\mathbf{F}\|_{F}^{2} \leq P_t$}, \label{eq:power} \\
    & \scalebox{0.9}{$|\boldsymbol{\Phi}(i,i)| = 1, \quad \forall i$}, \label{eq:phi} \\
    & \scalebox{0.9}{$\mathbf{A} = \mathbf{A}_{a} \mathbf{A}_{a}^{T}$}, \label{eq:A-Aa} \\
    & \scalebox{0.9}{$\mathbf{A}(i,i) \in \{0, 1\}, \quad i = 1, \dots, N$}, \label{eq:A} \\
    & \scalebox{0.9}{$\operatorname{tr}(\mathbf{A}) = a$}, \label{eq:a}
\end{align}
\end{subequations}
where $\mathbf{r} = [\mathbf{r}_1^T, \mathbf{r}_2^T]^T$ and $\mathbf{F} = [\mathbf{F}_1^T, \mathbf{F}_2^T]^T$.
In optimization problem (P1), constraints \eqref{eq:r1-vec} and \eqref{eq:r2-vec} represent the vectorization of the submatrices of $\mathbf{F}$. Constraint \eqref{eq:sinr} ensures that each user $k$ achieves a minimum SINR threshold $\overline{\gamma}_{k}$. Constraint \eqref{eq:power} imposes a total power budget $P_t$, while \eqref{eq:phi} enforces unit-modulus reflection coefficients. Constraints \eqref{eq:A-Aa}, \eqref{eq:A}, and \eqref{eq:a} define the feasible set for the mode selection matrices.

Problem (P1) is non-convex and challenging. To address this, we employ an AO framework that decomposes the problem into tractable sub-problems solved iteratively, integrating MM and penalty techniques. The procedure alternates between optimizing the beamforming matrix $\mathbf{F}$ with fixed $\mathbf{A}, \mathbf{A}_{a}$, and $\boldsymbol{\Phi}$, and optimizing the reflection coefficients $\boldsymbol{\Phi}$ along with the mode selection matrices $\mathbf{A}$ and $\mathbf{A}_{a}$ with fixed $\mathbf{F}$.

\subsection{Transmit Beamforming Matrix Optimization}

This section presents the proposed methodology for the joint design of beamforming, phase, and mode selection in the RDARS-aided ISAC system.

For fixed $\boldsymbol{\Phi}$, $\mathbf{A}$, and $\mathbf{A}_{a}$, the optimization problem with respect to $\mathbf{F}$ is given by
\begin{subequations}
\begin{align}
    \text{(P2)} \quad & \min_{\mathbf{F},\mathbf{r}} \quad  \frac{\mathbf{r}^H \mathbf{m} + \mathbf{m}^H \mathbf{r}}{\mathbf{r}^H \tilde{\mathbf{B}} \mathbf{r}} \nonumber\\
    \text{s.t.} \quad & \eqref{eq:r1-vec}, \eqref{eq:r2-vec}, \eqref{eq:sinr}, \eqref{eq:power}.
\end{align}
\end{subequations}

Let us define $\mathbf{F}_{1,k} \triangleq \mathbf{f}_{1,k} \mathbf{f}_{1,k}^H$, $\mathbf{F}_{2,k} \triangleq \mathbf{f}_{2,k} \mathbf{f}_{2,k}^H$, $\mathbf{F}_{12,k} \triangleq \mathbf{f}_{1,k} \mathbf{f}_{2,k}^H$, and $\mathbf{H}_{k} \triangleq \mathbf{h}_{k}^{*} \mathbf{h}_{k}^T$. Thus, the beamforming optimization problem (P2) can be reformulated as
\begin{subequations}
\begin{IEEEeqnarray}{l}
    \text{(P3)} \quad \min_{\substack{\mathbf{r}_1,\mathbf{r}_2, {(\mathbf{F}_{1,k})}_{k=1}^K, \\ {(\mathbf{F}_{2,k})}_{k=1}^K, {(\mathbf{F}_{12,k})}_{k=1}^K}} 
    \scalebox{0.9}{$\frac{\mathbf{r}^H \mathbf{m} + \mathbf{m}^H \mathbf{r}}{\mathbf{r}^H \tilde{\mathbf{B}} \mathbf{r}}$} \nonumber\\
    \text{s.t.} \quad \quad \quad \eqref{eq:r1-vec}, \eqref{eq:r2-vec}, \IEEEyesnumber \\
    \scalebox{0.9}{$\sum_{k=1}^{K} \operatorname{tr} \left\{ \begin{bmatrix} \mathbf{F}_{1,k} & \mathbf{F}_{12,k} \\ \mathbf{F}_{12,k}^H & \mathbf{F}_{2,k} \end{bmatrix} \right\} \leq P_t$}, \label{eq:power-updated} \\
    \scalebox{0.72}{$\displaystyle \operatorname{tr} \left\{ \mathbf{H}_k \begin{bmatrix} \mathbf{F}_{1,k} & \mathbf{F}_{12,k} \\ \mathbf{F}_{12,k}^H & \mathbf{F}_{2,k} \end{bmatrix} \right\} - \overline{\gamma}_{k} \sum_{{\substack{{i=1}\\{i\neq k}}}}^{K} \operatorname{tr} \left\{ \mathbf{H}_k \begin{bmatrix} \mathbf{F}_{1,i} & \mathbf{F}_{12,i} \\ \mathbf{F}_{12,i}^H & \mathbf{F}_{2,i} \end{bmatrix} \right\} \geq \overline{\gamma}_{k} \sigma_{k}^2$}.
    \label{New-SINR} 
    \IEEEyesnumber
\end{IEEEeqnarray}
\end{subequations}
To solve the non-convex problem (P3), we combine Dinkelbach's method \cite{dinkelbach1967} with a successive convex approximation of the lifted transmit-covariance problem. Define
\begin{equation}
\label{eq:Xk_definition}
\mathbf{X}_k \triangleq \begin{bmatrix}
\mathbf{F}_{1,k} & \mathbf{F}_{12,k}\\
\mathbf{F}_{12,k}^H & \mathbf{F}_{2,k}
\end{bmatrix}\succeq\mathbf0,\qquad \operatorname{rank}(\mathbf X_k)=1,\quad \forall k.
\end{equation}
Then,
\begin{equation}
\label{eq:r1r2_Xk}
\begin{aligned}
\mathbf r_1&=\operatorname{vec}\!\left(\sum_{k=1}^K\mathbf X_k^{(1,1)}\right),\quad
\mathbf r_2=\operatorname{vec}\!\left(\sum_{k=1}^K\mathbf X_k^{(2,2)}\right),\\
\mathbf r&=[\mathbf r_1^T,\mathbf r_2^T]^T.
\end{aligned}
\end{equation}
where $\mathbf X_k^{(i,j)}$ denotes the $(i,j)$-th block. Using Appendix~A, write
\begin{equation}
\label{eq:frac_obj}
f_0(\mathbf r)=\frac{N(\mathbf r)}{D(\mathbf r)},\qquad
N(\mathbf r)=\mathbf r^H\mathbf m+\mathbf m^H\mathbf r,\quad
D(\mathbf r)=\mathbf r^H\widetilde{\mathbf B}\mathbf r.
\end{equation}
On the identifiable feasible set, $D(\mathbf r)>0$. The matrix $\widetilde{\mathbf B}$ is Hermitian but is not assumed positive semidefinite. At Dinkelbach iteration $i$, we therefore solve a local convex approximation of
\begin{equation}
\label{eq:dinkelbach_sub}
\min_{\mathbf r_1,\mathbf r_2,\{\mathbf X_k\}}\;N(\mathbf r)-\varsigma^{(i)}D(\mathbf r),
\end{equation}
and update
\begin{equation}
\label{eq:varsigma_update}
\varsigma^{(i+1)}=\frac{N(\mathbf r^{(i+1)})}{D(\mathbf r^{(i+1)})}.
\end{equation}
Let $\widetilde{\mathbf B}=\mathbf B_+-\mathbf B_-$ be its spectral decomposition into positive and negative semidefinite parts, with $\mathbf B_+,\mathbf B_-\succeq\mathbf0$. At the current iterate $\mathbf r^{(t)}$, the concave term is majorized as
\begin{equation}
-\varsigma^{(i)}\mathbf r^H\mathbf B_+\mathbf r
\le -2\varsigma^{(i)}\Re\{\mathbf r^H\mathbf B_+\mathbf r^{(t)}\}+\mathrm{const}.
\label{eq:D_lower_bound}
\end{equation}
After dropping constants, the resulting convex subproblem is
\begin{subequations}
\label{eq:convex_P4}
\begin{IEEEeqnarray}{l}
\text{(P4)}\quad \min_{\mathbf r_1,\mathbf r_2,\{\mathbf X_k\}}
\;2\Re\{\mathbf m^H\mathbf r\}
-2\varsigma^{(i)}\Re\{\mathbf r^H\mathbf B_+\mathbf r^{(t)}\}\nonumber\\
\hspace{24mm}+\varsigma^{(i)}\mathbf r^H\mathbf B_-\mathbf r \nonumber\\
\text{s.t.}\quad \eqref{eq:r1r2_Xk},\;\eqref{New-SINR},\quad
\mathbf X_k\succeq\mathbf0,\ \forall k, \nonumber\\
\hspace{12mm}\sum_{k=1}^K\operatorname{tr}(\mathbf X_k)\le P_t.
\end{IEEEeqnarray}
\end{subequations}
Problem (P4) is solved with a standard convex solver such as CVX \cite{cvx2016} after dropping the rank-one constraints in \eqref{eq:Xk_definition}; hence, it is an SDR of (P3). If the relaxed solution is rank one, the beamformers follow directly by factorization. Otherwise, a standard Gaussian-randomization/rank-one recovery followed by SINR-feasibility power rescaling is used, and the feasible candidate with the smallest original CRB is retained. Accordingly, this transmit update is relaxation based and no global-optimality claim is made.

\subsection{Reflection Matrix and Selection Matrices Optimization}

Variables \(\boldsymbol{\Phi}\), \(\mathbf{A}\) and \(\mathbf{A}_{a}\) share diagonal structures with unit-modulus properties: \((\text{diag}(\boldsymbol{\Phi}))^{H}\text{diag}(\boldsymbol{\Phi}) = N\), \((\text{diag}(\mathbf{A}))^{T}\text{diag}(\mathbf{A}) = a\), and \((\text{vec}(\mathbf{A}_{a}))^{T}\text{vec}(\mathbf{A}_{a}) = a\). Thus, we optimize them jointly. Given fixed \(\mathbf{F}_1\) and \(\mathbf{F}_2\), the corresponding sub-problem is given by:
\begin{subequations}
\begin{align}
    \text{(P5)} \quad & \min_{\boldsymbol{\Phi},\mathbf{A},\mathbf{A}_{a}} \quad \text{CRB} \nonumber \\
    & \text{s.t.} \quad \eqref{eq:sinr}, \eqref{eq:phi}, \eqref{eq:A-Aa}, \eqref{eq:A}, \eqref{eq:a}.
\end{align}
\end{subequations}

To handle the coupled constraints \eqref{eq:sinr} and \eqref{eq:A-Aa}, we employ the penalty method. Introduce auxiliary variables \(s_{k,i}\) such that:
\begin{align}
    & \scalebox{0.9}{$\frac{|s_{k,k}|^{2}}{\sum_{i \neq k}^{K} |s_{k,i}|^{2} + \sigma_{k}^{2}} \geq \overline{\gamma}_{k}$}, \quad \forall k = 1,\dots,K. \label{eq:Sk1} \\
    & \scalebox{0.9}{$\mathbf{h}_{k}^{T} \mathbf{f}_{i} = s_{k,i}$}, \quad \forall k, \forall i. \label{eq:Sk2}
\end{align}

Incorporating \eqref{eq:A-Aa} and \eqref{eq:Sk2} as penalty terms yields:
\begin{subequations}
\begin{align}
    \text{(P6)} \quad & \min_{\boldsymbol{\Phi},\mathbf{A},\mathbf{A}_{a},s_{k,i}} \quad \scalebox{0.9}{$\text{CRB} +\frac{1}{2\rho_{1}} \sum_{k=1}^{K} \sum_{i=1}^{K} |\mathbf{h}_{k}^{T} \mathbf{f}_{i} - s_{k,i}|^{2}$} \nonumber \\
    & \quad \quad \quad \quad \scalebox{0.9}{$+ \frac{1}{2\rho_{2}} \|\mathbf{A} - \mathbf{A}_{a} \mathbf{A}_{a}^{T}\|_{F}^{2}$} \label{eq:20a} \nonumber \\
    & \text{s.t.}
    \quad \eqref{eq:phi}, \eqref{eq:A},\eqref{eq:a},  \eqref{eq:Sk1}, 
\end{align}
\end{subequations}
where $\rho_{1}$ and $\rho_{2}$ are penalty coefficients. Problem (P6) is still non-convex, but the constraints are now decoupled across variables. To solve Problem (P6), we iteratively update the following four blocks.

\subsubsection{Update \(s_{k,i}\)}
\label{subsubsec:update_s}

With fixed \(\boldsymbol{\Phi}\), \(\mathbf{A}\), and \(\mathbf{A}_{a}\), the sub-problem for \(s_{k,i}\) is given by:
\begin{subequations}
\begin{align}
    \text{(P7)} \quad & \min_{s_{k,i}} \quad \scalebox{0.9}{$\sum_{k=1}^{K} \sum_{i=1}^{K} |\mathbf{h}_{k}^{T} \mathbf{f}_{i} - s_{k,i}|^{2}$} \nonumber \\
    & \text{s.t.} \quad \eqref{eq:Sk1}.
\end{align}
\end{subequations}

The objective consists of \(K\) independent terms, and for each \(k\) the variables \(\{s_{k,i}, \forall i\}\) are separable. Hence, \(K\) sub-problems can be solved independently. The \(k\)-th sub-problem is:
\begin{subequations}
\begin{align}
    \text{(P8)} \quad & \min_{s_{k,i}} \quad \scalebox{0.9}{$\sum_{i=1}^{K} |\mathbf{h}_{k}^{T} \mathbf{f}_{i} - s_{k,i}|^{2}$} \label{eq:23a} \nonumber \\
    & \text{s.t.}  \quad \eqref{eq:Sk1}.
\end{align}
\end{subequations}

Problem (P8) is a quadratically constrained quadratic program. Strong duality holds when Slater's condition is satisfied, enabling the use of the Lagrangian dual method \cite{boyd2004convex}. Introducing the dual variable \(\mu_k \ge 0\), the Lagrangian is:
\begin{align}
    \mathcal{L}(s_{k,i},\mu_k) = & \scalebox{0.9}{$\sum_{i=1}^K |\mathbf{h}_{k}^T \mathbf{f}_i - s_{k,i}|^2$} \nonumber \\
    & \scalebox{0.9}{$+ \mu_k \left( \overline{\gamma}_k \left( \sum_{i \neq k}^{K} |s_{k,i}|^2 + \sigma_{k}^2 \right) - |s_{k,k}|^2 \right)$}. \label{eq:lagrangian}
\end{align}

The dual function is:
\begin{equation}
\scalebox{0.9}{$f_{\text{dual}}(\mu_k) = \min_{s_{k,i}} L(s_{k,i},\mu_k)$}.
\end{equation}
For \(f_{\text{dual}}(\mu_k)\) to be bounded, \(\mu_k\) must satisfy \(0 \le \mu_k < 1\). The proof is omitted for simplicity. Setting the first-order derivative to zero yields the optimal value of $s_{k,i}$ is:
\begin{equation}
s_{k,i}^{\star}=
\scalebox{0.9}{$\begin{cases}
\dfrac{\mathbf{h}_{k}^{T} \mathbf{f}_i}{1+\mu_k \overline{\gamma}_k}, & i\neq k,\\[0.3em]
\dfrac{\mathbf{h}_{k}^{T} \mathbf{f}_k}{1-\mu_k}, & i=k,
\end{cases}$}
\label{eq:opt_ski}
\end{equation}
where \(\mu_k\) is determined to satisfy the complementary slackness condition:
\begin{equation}
\scalebox{0.9}{$\mu_k\!\left( \overline{\gamma}_k\!\left(\sum_{i\neq k} |s_{k,i}^{\star}(\mu_k)|^2 + \sigma_{k}^2 \right) - |s_{k,k}^{\star}(\mu_k)|^2 \right)=0$}.
\end{equation}
This can be efficiently found via a one-dimensional search technique.

\subsubsection{Update $\boldsymbol{\phi}$}
\label{subsubsec:update_phi}
With $s_{k,i}$, $\mathbf A$, and $\mathbf A_a$ fixed, we use the equality penalty already introduced in (P6), rather than introducing a second SINR-violation penalty. Partition $\mathbf f_i=[\mathbf f_{1,i}^T,\mathbf f_{2,i}^T]^T$ and define
\begin{align}
c_{k,i}&\triangleq\mathbf h_{bu,k}^T\mathbf f_{1,i}+\mathbf h_{ru,k}^T\mathbf A_a\mathbf f_{2,i},\\
\mathbf b_{k,i}&\triangleq\operatorname{diag}(\mathbf H_{br}\mathbf f_{1,i})(\mathbf I-\mathbf A)\mathbf h_{ru,k}.
\end{align}
Then $\mathbf h_k^T\mathbf f_i=c_{k,i}+\boldsymbol\phi^T\mathbf b_{k,i}$, and the phase-update subproblem is
\begin{subequations}
\begin{align}
\text{(P9)}\quad \min_{\boldsymbol\phi}\;&\operatorname{CRB}(\boldsymbol\phi)+\frac{1}{2\rho_1}\sum_{k=1}^{K}\sum_{i=1}^{K}|\boldsymbol\phi^T\mathbf b_{k,i}+c_{k,i}-s_{k,i}|^2\nonumber\\
\text{s.t.}\;&|\phi_n|=1,\quad \forall n.
\end{align}
\end{subequations}
For fixed $\mathbf A$ and $\mathbf A_a$, only $\mathbf G_3$ depends on $\boldsymbol\phi$. Because both $g$ and its $\theta_2$-derivative enter the corrected FIM, Appendix~B writes $\operatorname{CRB}(\boldsymbol\phi)=F(S_0(\boldsymbol\phi),S_1(\boldsymbol\phi))$, where $S_0=|g|^2=\boldsymbol\phi^H\mathbf Q_0\boldsymbol\phi$ and $S_1=|\partial g/\partial\theta_2|^2=\boldsymbol\phi^H\mathbf Q_1\boldsymbol\phi$. Moreover, the communication penalty can be written as $\boldsymbol\phi^H\mathbf Q_c\boldsymbol\phi+2\Re\{\boldsymbol\phi^H\mathbf q_c\}+\mathrm{const}$, where
\begin{equation}
\mathbf Q_c=\sum_{k,i}\mathbf b_{k,i}^{*}\mathbf b_{k,i}^{T},\qquad
\mathbf q_c=\sum_{k,i}\mathbf b_{k,i}^{*}(c_{k,i}-s_{k,i}).
\end{equation}
At iteration $t$, we use the local first-order SCA model of $F(S_0,S_1)$ at $(s_{0,t},s_{1,t})$, with $\mu_{q,t}=\partial F/\partial S_q$, $q\in\{0,1\}$. This gives the local quadratic problem
\begin{subequations}
\begin{align}
\text{(P10)}\quad \min_{\boldsymbol\phi}\;&\boldsymbol\phi^H\mathbf S_t\boldsymbol\phi+2\Re\{\boldsymbol\phi^H\mathbf v_t\}\nonumber\\
\text{s.t.}\;&|\phi_n|=1,\quad \forall n,
\end{align}
\end{subequations}
where $\mathbf S_t=\mu_{0,t}\mathbf Q_0+\mu_{1,t}\mathbf Q_1+(2\rho_1)^{-1}\mathbf Q_c$ and $\mathbf v_t=(2\rho_1)^{-1}\mathbf q_c$. With $\mathbf M_t=\lambda_{\max}(\mathbf S_t)\mathbf I$, unimodular quadratic majorization \cite{Sun2017} yields the candidate
\begin{equation}
\boldsymbol\phi^{\rm c}=-\exp\!\left(j\arg\left((\mathbf S_t-\mathbf M_t)\boldsymbol\phi^{(t)}+\mathbf v_t\right)\right).
\label{eq:phi-opt}
\end{equation}
Since the first-order CRB model is local rather than a global upper bound, the candidate is accepted only if it does not increase the original fixed-$\rho_1$ objective in (P9); otherwise the current phase vector is retained.

\subsubsection{Update of $\mathbf a$}
Let $\mathbf a=[a_1,\ldots,a_N]^T$ and $\mathbf A=\operatorname{diag}(\mathbf a)$. With $\boldsymbol\phi$, $s_{k,i}$, and $\mathbf A_a$ fixed, the block subproblem consistent with (P6) is
\begin{subequations}
\begin{align}
\text{(P11)}\quad \min_{\mathbf a}\;&\operatorname{CRB}(\mathbf a)+\frac{1}{2\rho_1}\sum_{k,i}|\mathbf h_k^T\mathbf f_i-s_{k,i}|^2\nonumber\\
&+\frac{1}{2\rho_2}\|\mathbf A-\mathbf A_a\mathbf A_a^T\|_F^2\nonumber\\
\text{s.t.}\;&a_n\in\{0,1\},\quad \mathbf1^T\mathbf a=a.
\end{align}
\end{subequations}
For Path~3, define $g_0\triangleq\mathbf c^H(\theta_4)\boldsymbol\Phi\mathbf c(\theta_5)$ and $\mathbf v\triangleq\overline{\mathbf c(\theta_4)}\odot\boldsymbol\phi\odot\mathbf c(\theta_5)$, so that $g=g_0-\mathbf a^T\mathbf v$. Likewise, with $\dot{\mathbf c}(\theta_5)=\partial\mathbf c(\theta_5)/\partial\theta_2$, we have $\partial g/\partial\theta_2=\dot g_0-\mathbf a^T\dot{\mathbf v}$, where $\dot g_0\triangleq\mathbf c^H(\theta_4)\boldsymbol\Phi\dot{\mathbf c}(\theta_5)$ and $\dot{\mathbf v}\triangleq\overline{\mathbf c(\theta_4)}\odot\boldsymbol\phi\odot\dot{\mathbf c}(\theta_5)$. Hence the Path-3 information is quadratic in the binary vector $\mathbf a$.
For the communication penalty, define
\begin{align}
\mathbf t_{k,i}&\triangleq-\operatorname{diag}(\mathbf h_{ru,k})\boldsymbol\Phi\mathbf H_{br}\mathbf f_{1,i},\\
d_{k,i}&\triangleq\mathbf h_{bu,k}^T\mathbf f_{1,i}+\mathbf h_{ru,k}^T\boldsymbol\Phi\mathbf H_{br}\mathbf f_{1,i}+\mathbf h_{ru,k}^T\mathbf A_a\mathbf f_{2,i},
\end{align}
so that $\mathbf h_k^T\mathbf f_i=\mathbf a^T\mathbf t_{k,i}+d_{k,i}$. After the same local first-order approximation of the CRB, collect the resulting quadratic terms as
\begin{equation}
G(\mathbf a|\mathbf a^{(t)})=\mathbf a^T\mathbf S_t\mathbf a+2\Re\{\mathbf a^T\mathbf v_t\}+\mathrm{const}.
\label{G(a)}
\end{equation}
Because $\mathbf a^T\mathbf a=a$ is fixed, with $\lambda_t=\lambda_{\max}(\mathbf S_t)$ the quadratic term is majorized by
\begin{equation}
\mathbf a^T\mathbf S_t\mathbf a\le2\Re\{\mathbf a^T(\mathbf S_t-\lambda_t\mathbf I)\mathbf a^{(t)}\}+\mathrm{const}.
\label{eq:linear_majorant}
\end{equation}
Thus, defining $\mathbf e=2\Re\{(\mathbf S_t-\lambda_t\mathbf I)\mathbf a^{(t)}+\mathbf v_t\}$, the local binary problem is
\begin{subequations}
\begin{align}
\text{(P12)}\quad \min_{\mathbf a}\;&\mathbf a^T\mathbf e\nonumber\\
\text{s.t.}\;&a_n\in\{0,1\},\quad \mathbf1^T\mathbf a=a.
\end{align}
\end{subequations}
Its solution selects the $a$ smallest entries of $\mathbf e$. If $\mathcal L$ is their index set,
\begin{equation}
\mathbf a^{\rm c}(n)=\begin{cases}1,&n\in\mathcal L,\\0,&\text{otherwise}.
\end{cases}
\label{eq:A-opt}
\end{equation}
The candidate is accepted only if the original fixed-penalty objective decreases; otherwise the current selection is retained.

\begin{algorithm}
\caption{CRB Optimization for Joint Beamforming and Mode Selection in RDARS-aided ISAC.}
\label{alg:main}
\begin{algorithmic}[1]
\REQUIRE $\mathbf h_{bu,k}$, $\mathbf h_{ru,k}$, $P_t$, $\sigma_k^2$, $\overline\gamma_k$, $\rho_1$, $\rho_2$, and $0<c_1,c_2<1$.
\STATE Initialize $\boldsymbol\Phi^0$, $\mathbf F^0$, $\mathbf A^0$, and $\mathbf A_a^0$.
\REPEAT
\STATE Update $\mathbf F$ by solving relaxed (P4), with rank-one recovery if required.
\STATE Update $s_{k,i}$ based on \eqref{eq:opt_ski}.
\STATE Update $\boldsymbol\phi$ from \eqref{eq:phi-opt} with the fixed-penalty acceptance check.
\STATE Update $\mathbf A$ from \eqref{eq:A-opt} with the fixed-penalty acceptance check.
\STATE Update $\mathbf A_a$ by solving (P13).
\STATE $\rho_1\gets c_1\rho_1$, $\rho_2\gets c_2\rho_2$.
\UNTIL{the stopping criterion is satisfied.}
\ENSURE $\mathbf F^*$, $\boldsymbol\Phi^*$, $\mathbf A^*$, $\mathbf A_a^*$.
\end{algorithmic}
\end{algorithm}

\begin{figure*}[!t]
    \centering
   \begin{minipage}[t]{0.49\textwidth}
    \centering
    \centerline{\includegraphics[width=7cm,height=4.28cm]{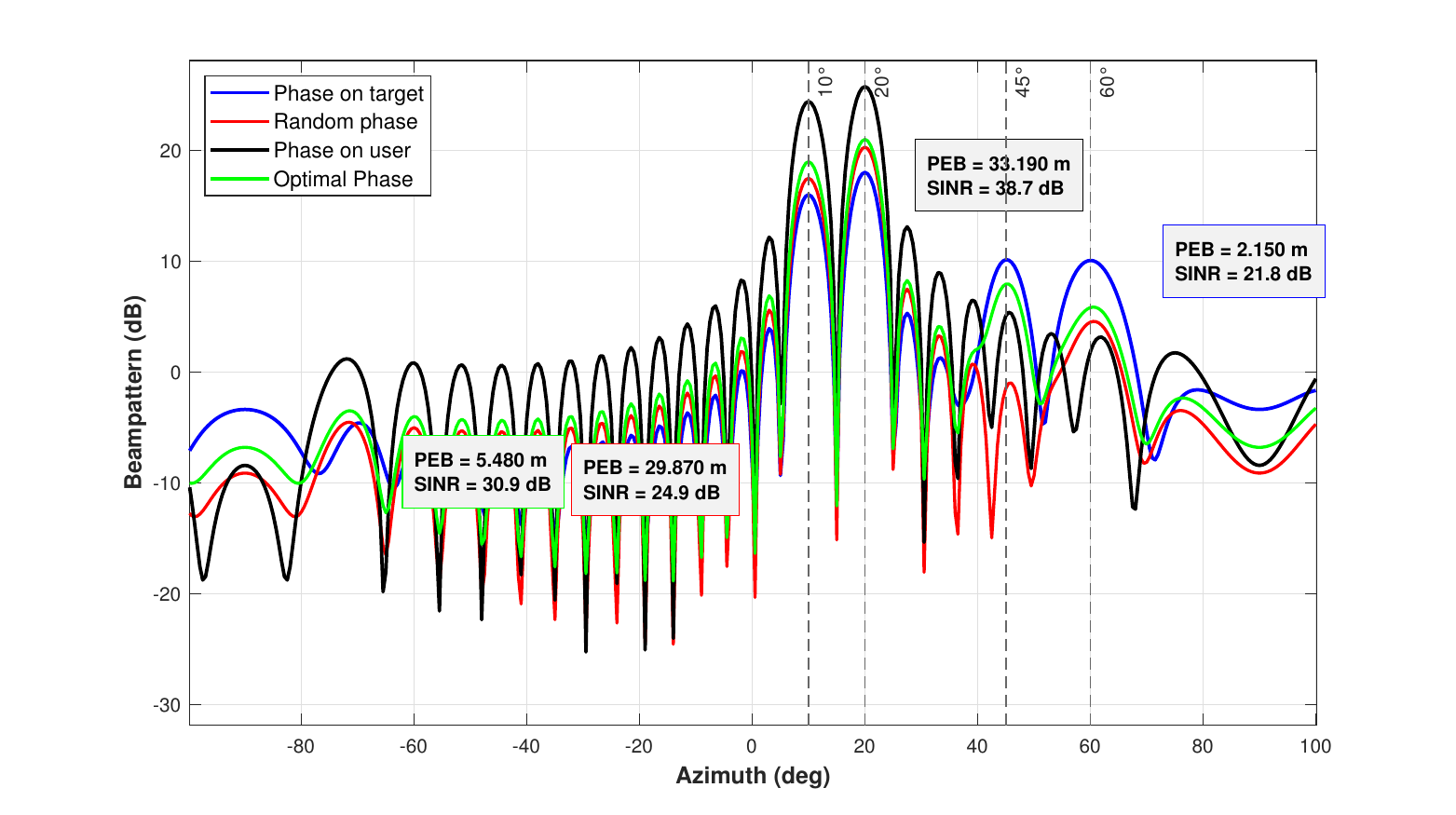}}
    \caption[The BS beampattern]{\small The BS beampattern, for the following RDARS phase profiles: reflection beam focused on target, on users, randomly generated and optimally designed.}
    \label{fig:BS-Beampattern}
\end{minipage}
    \hfill
    \begin{minipage}[t]{0.49\textwidth}
        \centering
        \centerline{\includegraphics[width=7cm,height=4.28cm]{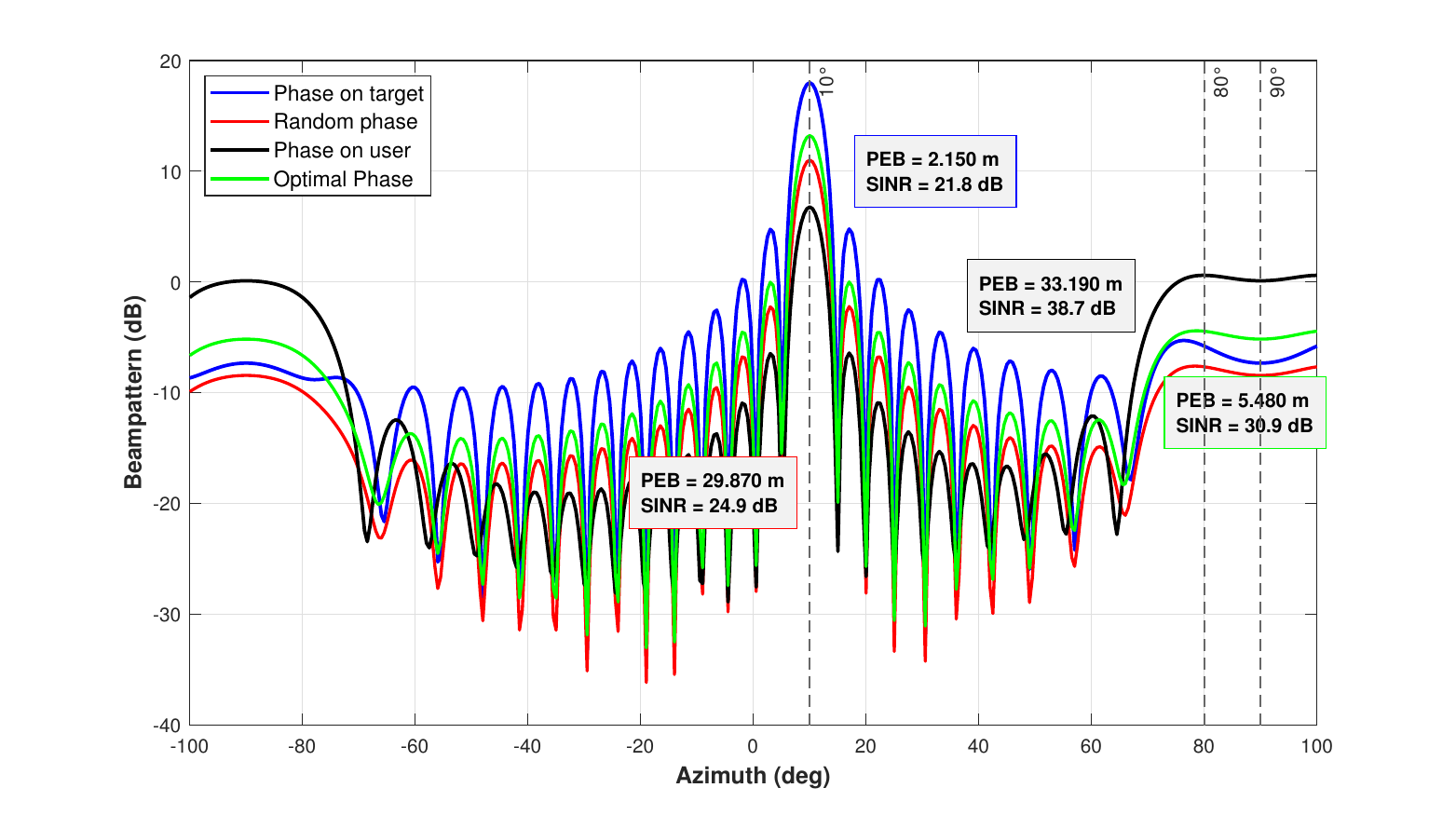
        }}
        \caption[The RDARS beampattern]{\small The beampattern of the RDARS connected elements for various phase profile configurations: reflection beam focused on target, on users, randomly generated and optimally designed.}
        \label{fig:RDARS-Beampattern}
    \end{minipage}
\end{figure*}

\subsubsection{Update of $\mathbf A_a$}
With $\boldsymbol\phi$, $s_{k,i}$, and $\mathbf A$ fixed, the connected-element selection matrix is updated from
\begin{subequations}
\begin{align}
\text{(P13)}\quad \min_{\mathbf A_a}\;&\operatorname{CRB}(\mathbf A_a)+\frac{1}{2\rho_1}\sum_{k,i}|\mathbf h_k^T\mathbf f_i-s_{k,i}|^2\nonumber\\
&+\frac{1}{2\rho_2}\|\mathbf A-\mathbf A_a\mathbf A_a^T\|_F^2\nonumber\\
\text{s.t.}\;&\sum_{l=1}^N[\mathbf A_a]_{l,j}=1,\quad [\mathbf A_a]_{l,j}\in\{0,1\},\nonumber\\
&\mathbf A_a^T\mathbf A_a=\mathbf I_a.
\end{align}
\end{subequations}
The last constraint ensures that the connected elements are distinct. Using a first-order local model of the CRB and the quadratic penalties, the resulting linearized binary subproblem is a finite assignment problem over $a$ columns and $N$ candidate rows; it is solved under the distinctness constraint, and the new $\mathbf A_a$ is accepted only if the original fixed-penalty objective is non-increasing. This update explicitly preserves the dependence of $\mathbf G_2$, $\mathbf G_4$, and $\mathbf G_5$ on the physical connected-element locations.

Problem (P1) is therefore handled by a local AO/penalty algorithm for a non-convex mixed-integer problem; global optimality is not claimed.

\subsection{Convergence and Complexity Analysis}
For fixed penalty parameters $\rho_1$ and $\rho_2$, the accepted $\boldsymbol\phi$, $\mathbf A$, and $\mathbf A_a$ updates are non-increasing in the penalized objective in (P6). The transmit block is solved through an SDR/SCA step, and a recovered beamformer is retained only when it is feasible and does not increase the fixed-RDARS CRB. Hence, within a fixed-penalty stage, the accepted objective sequence is bounded below and converges in value. Because (P1) contains rank-one, unit-modulus, and binary constraints, neither a KKT point nor a global optimum of the original mixed-integer problem is guaranteed. The continuation $\rho_1,\rho_2\downarrow0$ progressively enforces the coupling equalities; monotonicity is claimed only within each fixed-penalty stage.

Let $d=M+a$ and $n_x=\mathcal O(Kd^2)$ denote the number of real scalar variables in the lifted transmit block. A generic interior-point solution of (P4) has conservative worst-case scaling $\mathcal O(I_{\rm IP}n_x^3)=\mathcal O(I_{\rm IP}K^3d^6)$, where $I_{\rm IP}$ is the number of Newton steps; rank-one recovery adds at most $\mathcal O(Kd^3)$ per randomization batch. The phase update requires the dominant eigenvalue of an $N\times N$ Hermitian matrix, with $\mathcal O(N^3)$ complexity using a dense eigensolver. Updating $s_{k,i}$ requires $K$ one-dimensional searches with $K$ interference terms, i.e., $\mathcal O(K^2\log(1/\epsilon))$. The $\mathbf A$ update requires sorting, $\mathcal O(N\log N)$, while the linearized $\mathbf A_a$ assignment is polynomial in $a$ and $N$ and is negligible for $a\ll N$ compared with the SDP step. Thus, the lifted transmit update dominates the worst-case per-iteration complexity.

\section{Simulation Results} \label{sec:4}

Numerical simulations evaluate the proposed approach. The simulation parameters are listed in Table~\ref{tab:settings}.

\begin{table}[h]
\caption{The Main Simulation Parameters}
\centering
\begin{tabular}{|l|c|c|}  
\hline
\textbf{Description} & \textbf{Parameter} & \textbf{Value} \\
\hline
Number of transmit/receive antennas at BS & \(M\) & 16 \\
\hline
Number of RDARS elements & \(N\) & 120 \\
\hline
Number of connected elements & \(a\) & 3 \\
\hline 
Number of subcarriers & \(N_s\) & 1024\\
\hline
Total transmit power & \(P_t\) & 20 dBm \\
\hline
Noise power & \(\sigma_{k}^2\) & -80 dBm \\
\hline
Path loss exponent for channel \(H_{br}\) & \(\alpha_{br}\) & 2.4 \\
\hline
Users' SINR threshold & \(\overline{\gamma}_{k}\) & 15 dB \\
\hline
Carrier Frequency & \(f_c\) & 30 GHz \\
\hline
Bandwidth & \(B\) & 200 MHz \\
\hline
\end{tabular}
\label{tab:settings}
\end{table}

\begin{figure*}[!t]
    \centering
    \begin{minipage}[t]{0.49\textwidth}
        \centering
        \centerline{\includegraphics[width=7cm,height=4.28cm]{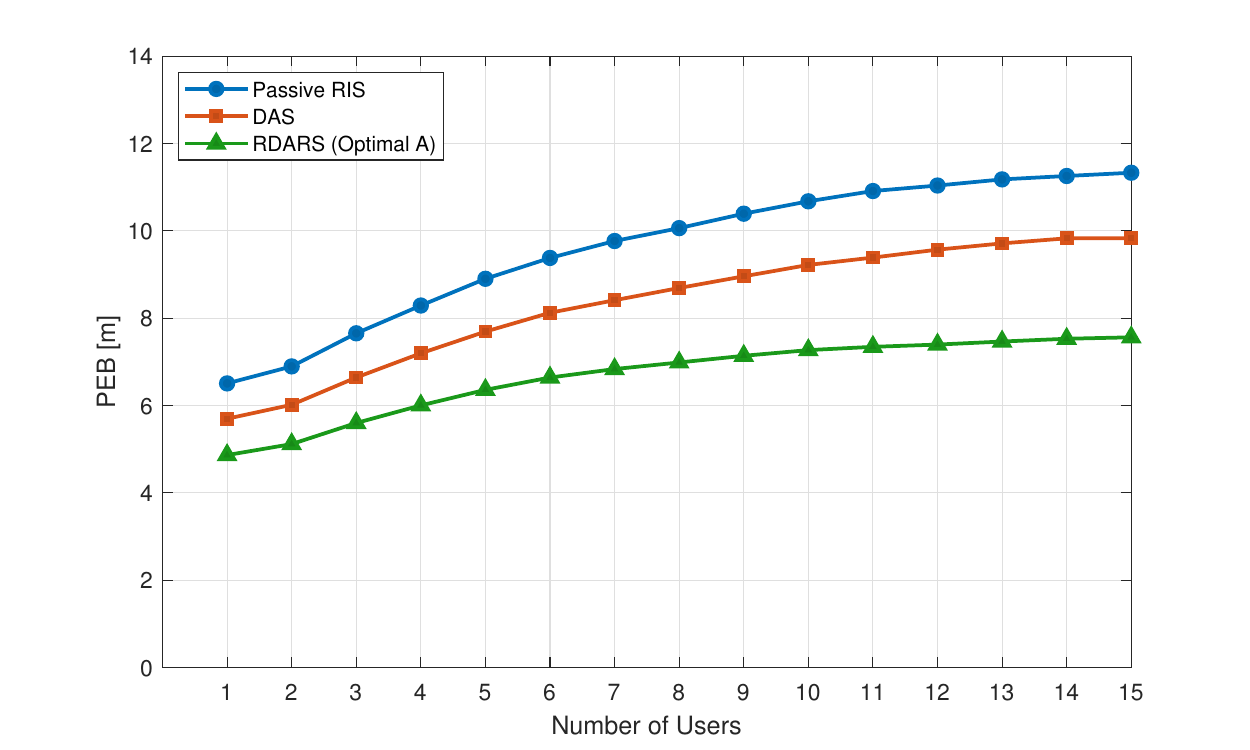}}
        \caption[The PEB vs. number of users]{\small The PEB as a function of  number of users, for the following structures: passive RIS, DAS and RDARS.}
        \label{fig:peb_nusers}
    \end{minipage}
    \hfill
    \begin{minipage}[t]{0.49\textwidth}
        \centering
        \centerline{\includegraphics[width=7cm,height=4.28cm]{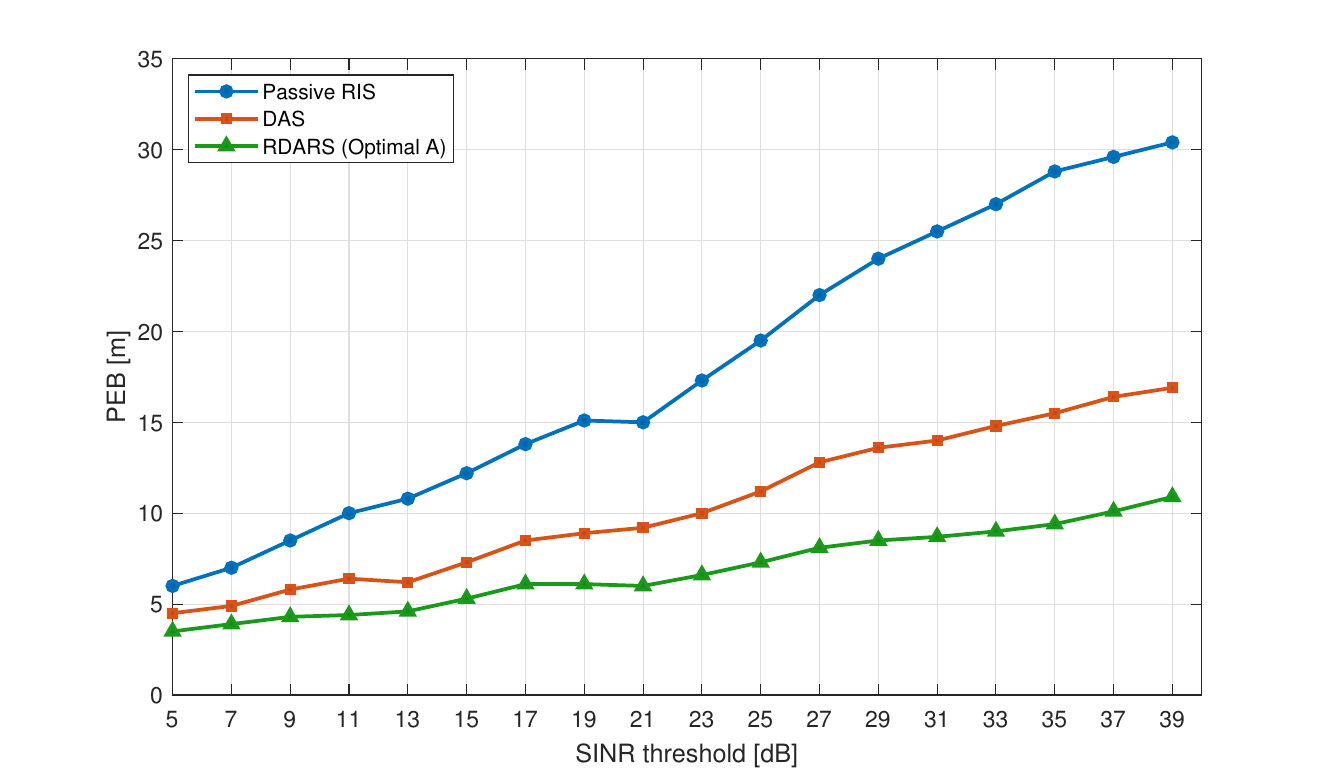}}
        \caption[PEB versus SINR threshold]{\small The PEB versus SINR threshold for different system architectures: passive RIS, DAS and RDARS.}
        \label{fig:peb_sinr_th}
    \end{minipage}
\end{figure*}
\begin{figure*}[!t]
    \centering
    \begin{minipage}[t]{0.49\textwidth}
        \centering
        \centerline{\includegraphics[width=7cm,height=4.28cm]{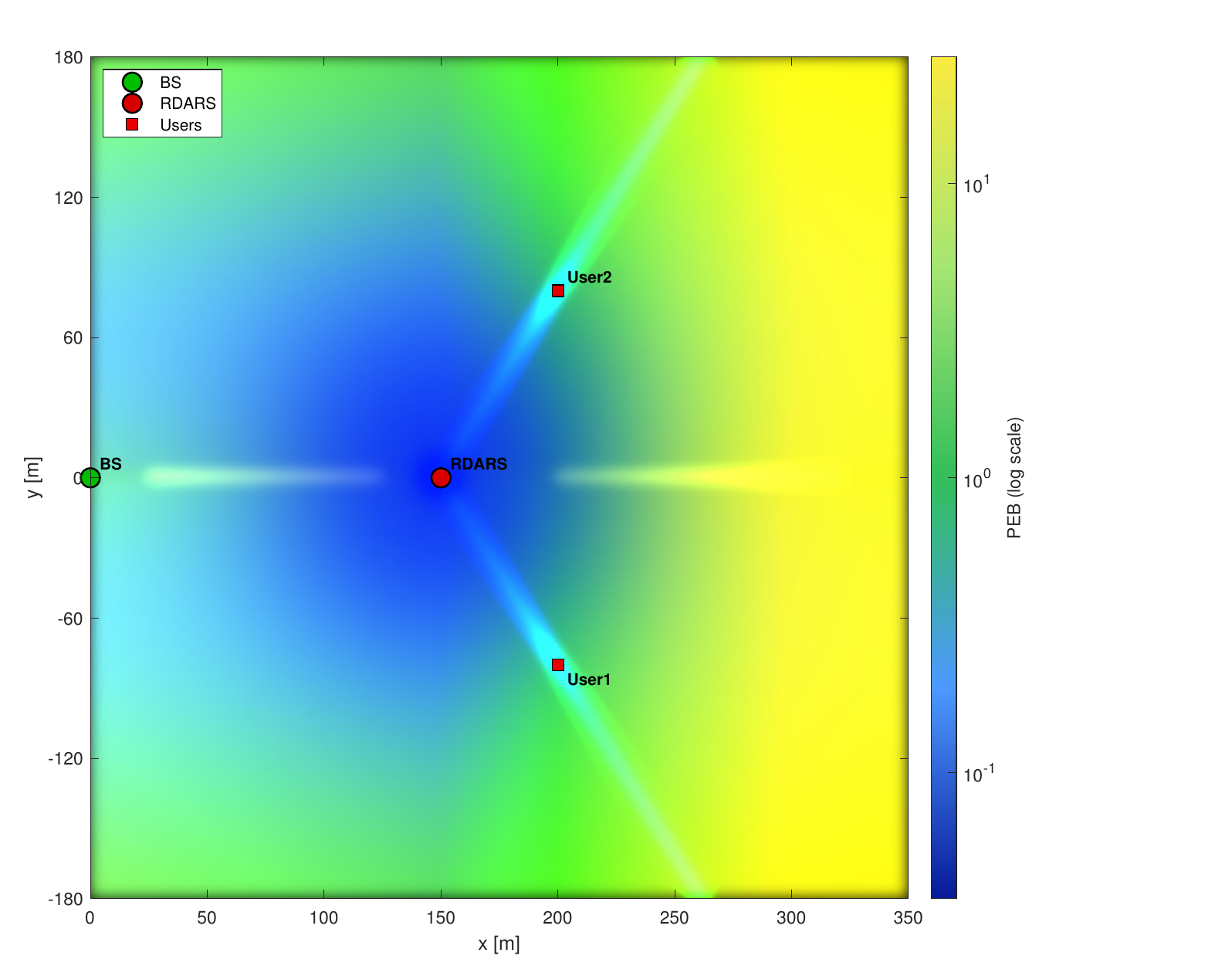}}
        \caption[Geometric Analysis]{\small The PEB for different positions of target, analysing the effect of geometry on ISAC performance.}
        \label{fig:geometry}
    \end{minipage}
    \hfill
    \begin{minipage}[t]{0.49\textwidth}
    \centering
    \centerline{\includegraphics[width=7cm,height=4.28cm]{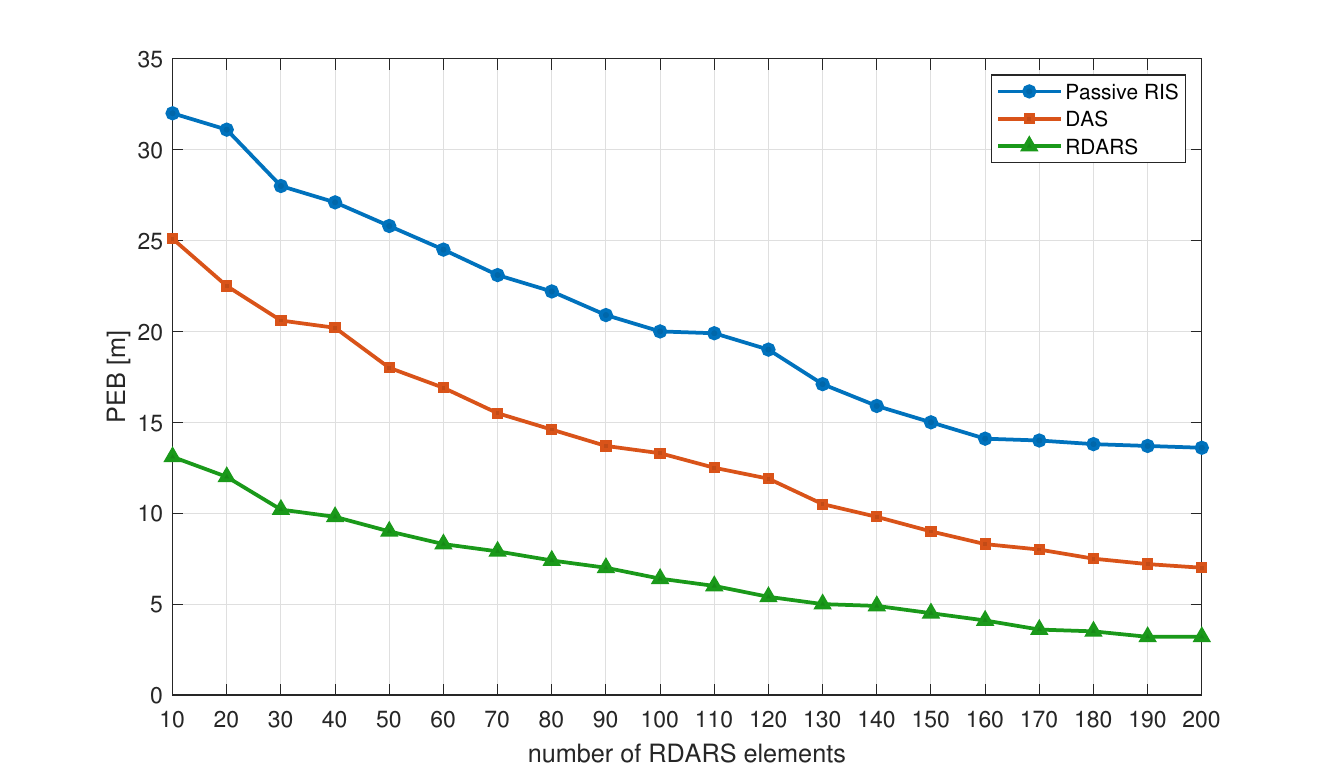}}
    \caption[PEB versus total number of RDARS elements]{\small The PEB as a function of total number of RDARS elements, for the following structures: passive RIS, DAS and RDARS.}
    \label{fig:peb_nrdars}
\end{minipage}
\end{figure*}

\subsection{Beampatterns}

This section presents the beampattern results for the BS and RDARS, shown in Fig.~\ref{fig:BS-Beampattern} and Fig.~\ref{fig:RDARS-Beampattern}. The scenario considers a target at \(60^\circ\) azimuth, two users at \(10^\circ\) and \(20^\circ\) and an RDARS at \(45^\circ\). The RDARS and target are located at a distance of 150 meters from BS. The RDARS phase have been adjusted in a way that four beam configurations are evaluated: reflection beam focused on target, on users, randomly generated and optimally designed. The figures show four distinct peaks near the users, RDARS and target.
At user's locations, the strongest peak corresponds to \textit{phase on user}, whereas the weakest corresponds to \textit{phase on target} case. Near the target location, however, this behavior is reversed. The \textit{optimal phase} produces an intermediate peak between these two extremes.
For performance comparison, the PEB is used as an accuracy metric that is equal to the square root of the CRB. Using the calculated FIM, the PEB is defined as follows:
\begin{equation}
\mathcal{P} \triangleq \sqrt{\operatorname{tr}\bigl(\mathbf{J}_e^{-1}(\mathbf{p})\bigr)} = \sqrt{ \operatorname{CRB}}.
\label{PEB_eq}
\end{equation}

Directing the beam closer to the users increases the communication power and improves the SINR performance, albeit at the cost of reduced radar power, leading to higher PEB and less accurate sensing. Conversely, steering the beam toward the target reverses this trade-off. To facilitate comparison across configurations, the corresponding SINR and PEB values are illustrated in Fig.~\ref{fig:BS-Beampattern}. The \textit{optimal phase case} achieves a balanced PEB of 5.48~m and outperforms other configurations in both metrics. Notably, while the "phase on target" configuration yields a lower PEB (2.150~m), it fails to meet the desired communication quality, achieving only 21.8~dB SINR compared to 30.9~dB for the optimal design, thereby demonstrating the effectiveness of our framework in balancing sensing and communication under constraints. To provide further engineering insights, we examine several special cases: the SISO case reveals the optimal phase alignment maximizing reflection gain; the single-user case illustrates the distribution gain of connected RDARS elements; the pure LoS scenario highlights the sensitivity of localization accuracy to the BS--RDARS--target geometry; and the high-SNR regime highlights the pronounced mode-selection gain, as the selection of connected elements shapes the FIM eigenvalue structure. Collectively, these simplified cases clarify the roles of reflection, distribution, and mode-selection gains, and offer practical design guidelines.

\subsection{ISAC Trade-off Analysis}

Fig.~\ref{fig:peb_nusers} and Fig.~\ref{fig:peb_sinr_th} investigate the trade-off between sensing and communication by analyzing the effects of two key parameters on the PEB: the number of users and the minimum SINR threshold \(\overline{\gamma}_{k}\). If the number of users or SINR threshold grow, the communication power will increase.
As a result, the radar power will decrease. therefore, PEB will increase, which indicates a decrease in localization accuracy.
The proposed RDARS framework generalizes two existing architectures: Passive RIS and DAS. These correspond to special cases of RDARS where the number of connected elements is zero or equal to the total number of elements, respectively. Since RDARS inherently combines both reflection and connection capabilities, it offers greater design flexibility and degrees of freedom. Also, due to the reduction in the number of connected elements in RDARS-based systems compared to DAS architecture, the need for costly RF chains is significantly reduced, which represents a very good advantage of RDARS. Consequently, as clearly illustrated in the results, RDARS achieves significantly better performance and a substantially lower PEB compared to passive RIS and DAS.

\begin{figure*}[!t]
\centering
\begin{minipage}[t]{0.49\textwidth}
    \centering
    \centerline{\includegraphics[width=7cm,height=4.28cm]{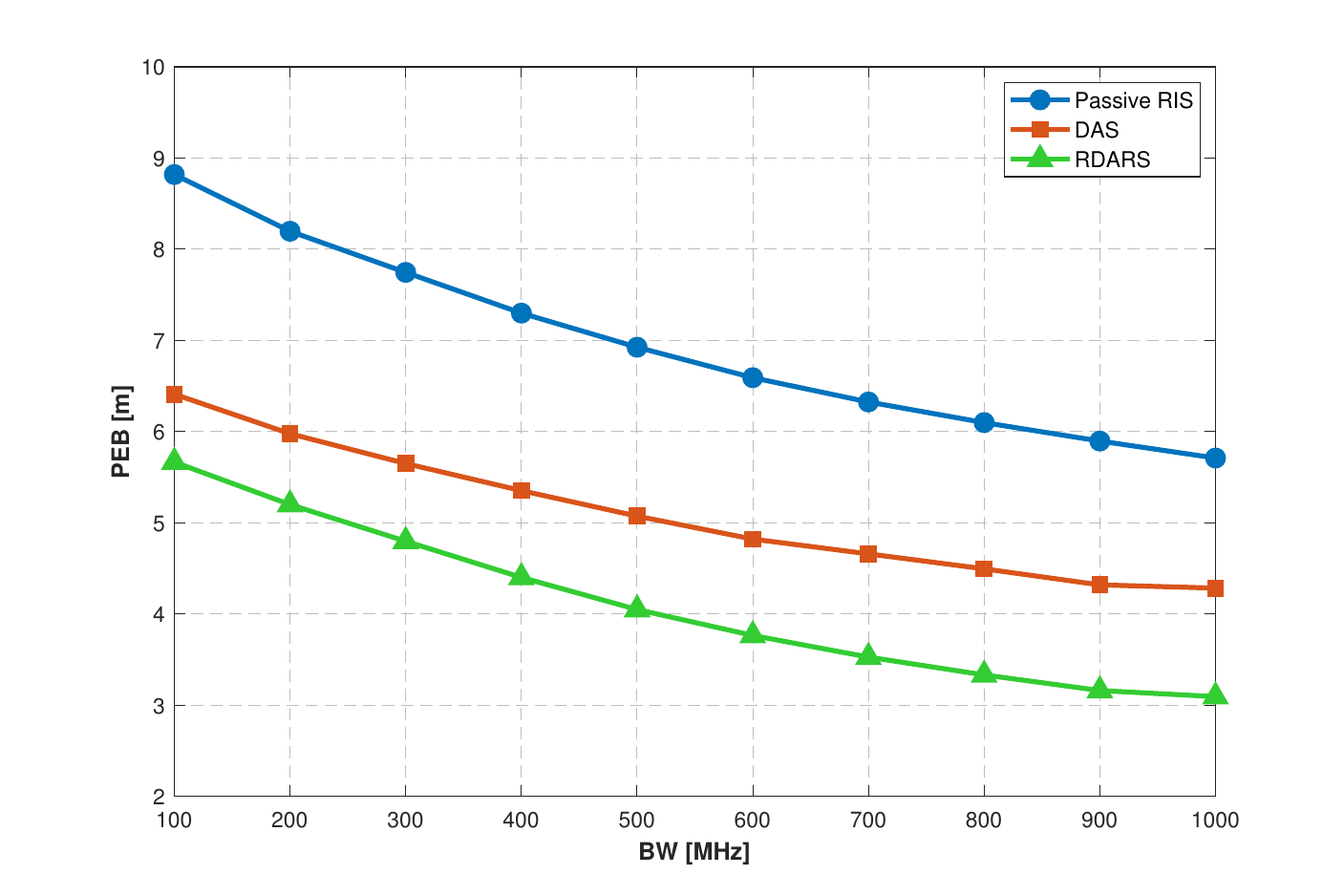}}
    \caption[PEB versus bandwidth]{\small The PEB as a function of signal bandwidth.}
    \label{fig:peb_bw}
\end{minipage}
\vspace{0.3cm}
\begin{minipage}[t]{0.49\textwidth}
    \centering
    \centerline{\includegraphics[width=7cm,height=4.28cm]{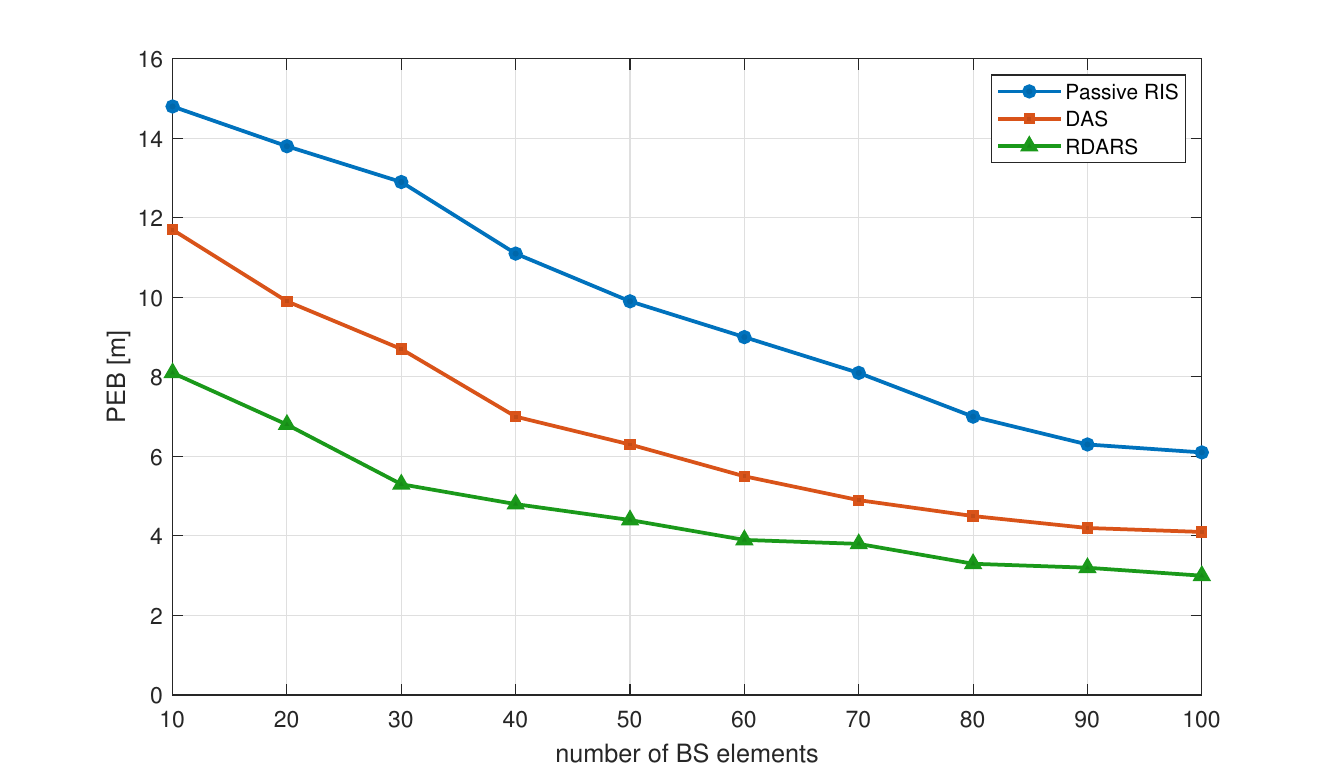}}
    \caption[PEB versus number of BS antennas]{\small The PEB versus number of BS antennas.}
    \label{fig:peb_nbs}
\end{minipage}
\hfill
\begin{minipage}[t]{0.49\textwidth}
    \centering
    \centerline{\includegraphics[width=7cm,height=4.28cm]{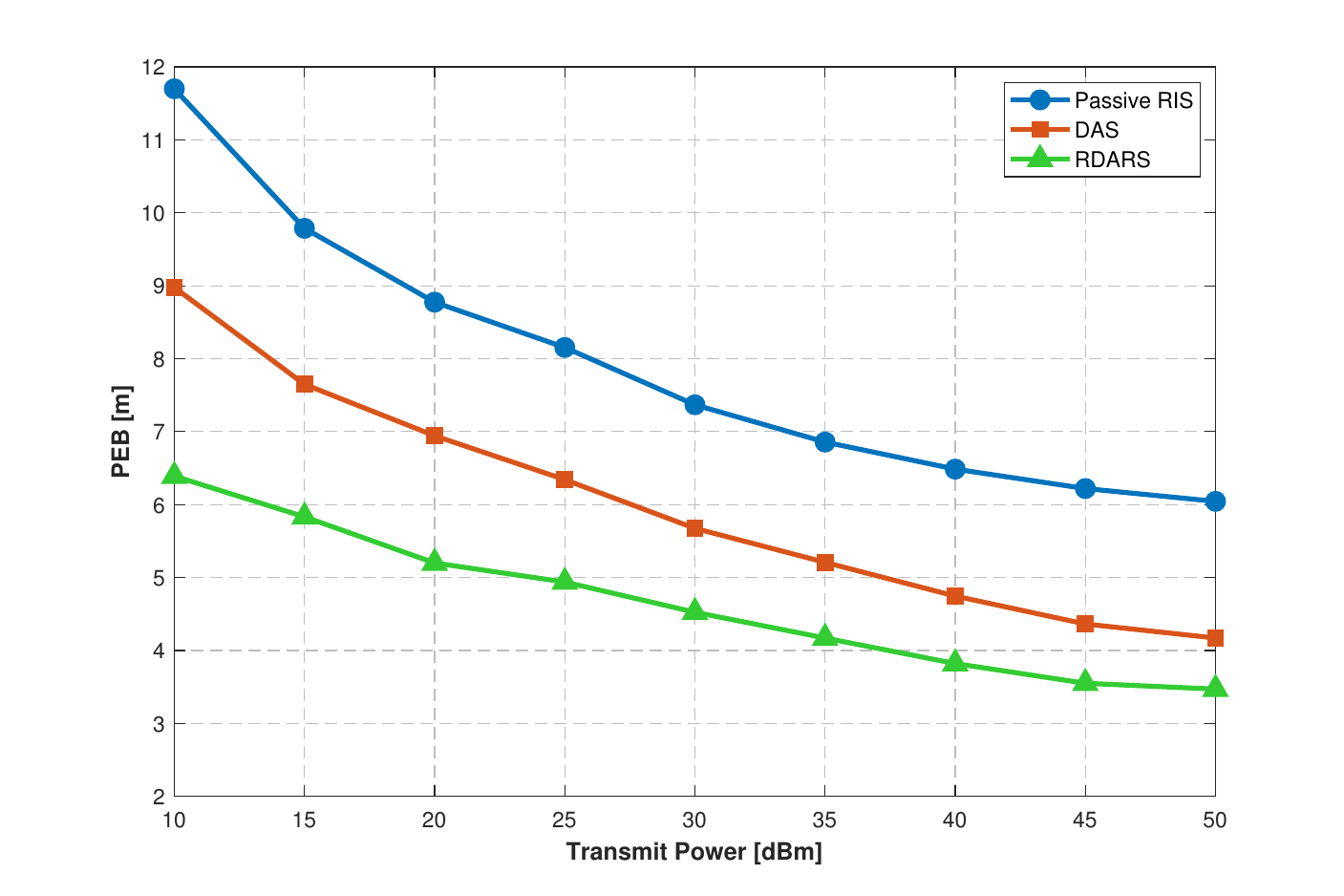}}
    \caption[PEB versus transmit power]{\small The PEB as a function of  transmit power, for the following structures: passive RIS, DAS and RDARS.}
    \label{fig:peb_pt}
\end{minipage}
\hfill
\begin{minipage}[t]{0.49\textwidth}
    \centering
    \centerline{\includegraphics[width=7cm,height=4.28cm]{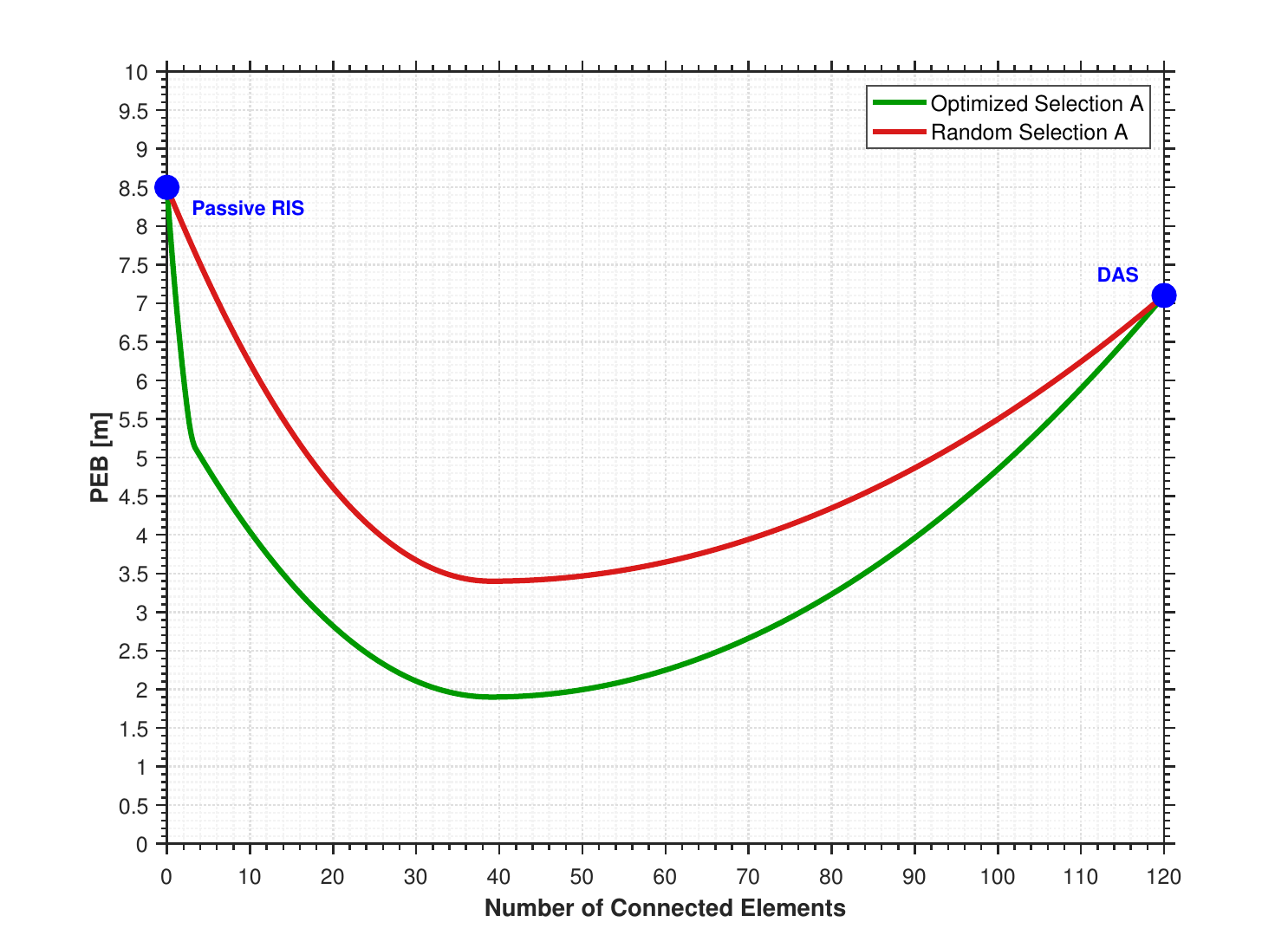}}
    \caption[PEB versus number of RDARS connected elements]{\small The PEB as a function of number of RDARS connected elements, analysing the selection gain of RDARS in comparison with passive RIS and DAS.}
    \label{fig:peb_pncon}
\end{minipage}
\end{figure*}

\subsection{Geometric Analysis}

The achievable localization accuracy is evaluated from a geometric perspective, shown in Fig.~\ref{fig:geometry}. The scenario consists of a BS at \((0,0)\), an RDARS at \((150,0)\) m, and two users at \((200, \pm 80)\) m. By varying the target location, the corresponding PEB is computed. The results, illustrated by the color spectrum in the contour plot, reveal several geometric insights. The region between the BS and the RDARS exhibits the lowest PEB, indicating the highest localization accuracy. As the target moves away from this baseline, the PEB gradually increases.
A high-error region with low accuracy is observed along the half-line extending from the RDARS location toward infinity. Along this half-line, the circular fixed-delay contours of the direct path are tangential to the elliptical fixed-delay contours of the indirect path, resulting in significantly increased sensitivity to noise in the localization process.
Furthermore, along the diagonal paths between the RDARS and the users, the PEB is relatively lower compared to the surrounding regions. This improvement can be attributed to enhanced signal illumination when the target, RDARS and users are approximately aligned.
These observations underscore the importance of geometric layout in ISAC system design.

\subsection{Analysis of System Design Parameters}

The final set of simulation results, presented in Figs.~\ref{fig:peb_nrdars}–\ref{fig:peb_pt}, analyzes key system parameters. As expected from communication-theoretic principles, the PEB consistently decreases with increasing bandwidth, number of BS antennas, total RDARS elements, and transmit power. Notably, even a small number of connected RDARS elements yields a substantially lower PEB compared to passive RIS and DAS, with the lowest PEB achieved at approximately 40 connected elements, although practical and economic constraints render such a large number cost-ineffective. The proposed RDARS framework inherently generalizes all benchmark architectures—passive RIS corresponds to $a = 0$, DAS corresponds to $a = N$. The results in Figs.~\ref{fig:peb_nrdars}, ~\ref{fig:peb_nbs}, and ~\ref{fig:peb_pncon} collectively confirm the scalability of the proposed framework to large-scale antenna systems, where the performance gains over passive RIS and DAS remain substantial as the number of BS antennas and RDARS elements increases, while the required number of active RF chains remains remarkably small. This hardware efficiency is particularly advantageous in massive MIMO deployments, where reducing active components is critical for cost and power consumption.
These parametric analyses provide valuable practical insights for future RDARS and ISAC system design.

\section{Conclusion} \label{sec:6}

This paper investigated a novel RDARS-aided ISAC framework for 6G networks, addressing the joint optimization of the base station's active beamforming and the RDARS's dynamic mode selection and passive reflection. The objective was to minimize the CRB for target localization accuracy while satisfying stringent communication quality-of-service constraints defined by user SINR thresholds. To tackle this challenging non-convex mixed-integer problem, we developed an efficient iterative algorithm based on the AO framework, effectively integrating MM and Penalty techniques. Extensive simulations demonstrated the superiority of the proposed design. By analyzing various beampatterns, system geometries and performance trade-offs, the results conclusively showed that the dynamically configured RDARS achieves a significantly lower PEB for a given SINR requirement compared to conventional passive RIS and DAS benchmarks. This superior performance stems from the synergistic exploitation of reflection, distribution and mode-selection gains unique to the RDARS architecture. Future research directions include extending this framework to multi-target localization scenarios, investigating RDARS-aided ISAC under imperfect channel state information, extending the model to multi-user localization downlink scenarios without ISAC, exploring the integration of RDARS with advanced waveform design for enhanced spectral efficiency, and extending the proposed framework to account for random and dynamic RCS fluctuations using Bayesian or modified Cramér–Rao bounds. Furthermore, a comprehensive investigation of multiple RDARS panels—encompassing optimal placement, inter-panel interference characterization, and trade-offs between the number of panels and the number of elements per panel—remains a compelling avenue for future work, as the present study establishes the foundational single-panel CRB-based optimization as a necessary first step toward such complex deployments.

\appendices
\section{Calculation of the FIM}
\label{app:A}

Based on the complex Gaussian distribution of the received signal in \eqref{eq:radar_signal}, the elements of $\mathbf{J}_{\boldsymbol{\eta}}$ are given by \cite{kay1993fundamentals}:
\begin{align}
\mathbf{J}_{\bm{\eta}} &= 
\scalebox{0.9}{$\begin{bmatrix} 
\mathbf{J}_{\bm{\tau},\bm{\tau}} & \mathbf{J}_{\bm{\tau},\bm{\theta}} & \mathbf{J}_{\bm{\tau},\bm{\alpha}} \\ 
\mathbf{J}_{\bm{\theta},\bm{\tau}} & \mathbf{J}_{\bm{\theta},\bm{\theta}} & \mathbf{J}_{\bm{\theta},\bm{\alpha}} \\ 
\mathbf{J}_{\bm{\alpha},\bm{\tau}} & \mathbf{J}_{\bm{\alpha},\bm{\theta}} & \mathbf{J}_{\bm{\alpha},\bm{\alpha}}
\end{bmatrix}$}, \\
\mathbf{J}^{(\textit{i},\textit{j})}_{\bm{\eta}} &= \scalebox{0.9}{$\frac{2}{\sigma^2}\sum_{n=-\frac{N_s-1}{2}}^{\frac{N_s-1}{2}} \Re\left\{
\frac{\partial \bm{\mu}^H[n]}{\partial \eta_{i,j}}
\frac{\partial \bm{\mu}[n]}{\partial \eta_{i,j}} \right\}$}. \label{S.B}
\end{align}
The FIM elements are derived from the partial derivatives of the signal model. The derivatives with respect to delays $\tau_i$ are:
\begin{align}
\frac{\partial \bm{\mu}[n]}{\partial \tau_{1}} &= 
\scalebox{0.9}{$\begin{bmatrix}
\alpha_{1}e^{-j2\pi n \Delta f\tau_{1}}[\mathbf{G}_{1},0]\mathbf{X}[n](-j2\pi n \Delta f) \\
\bm{0}
\end{bmatrix}$}, \label{p.d1}\\
\frac{\partial \bm{\mu}[n]}{\partial \tau_{2}} &= 
\scalebox{0.9}{$\begin{bmatrix}
\bm{0} \\
\alpha_{2}e^{-j2\pi n \Delta f\tau_{2}}[\mathbf{G}_{2},0]\mathbf{X}[n](-j2\pi n \Delta f)
\end{bmatrix}$},
\end{align}

Analogous expressions can be derived for $\tau_3$, $\tau_4$ and $\tau_5$. The derivatives with respect to delays angles $\theta_i$ and path losses $\alpha_i$ are given by:
\begin{align}
\frac{\partial \bm{\mu}[n]}{\partial \theta_{1}} &= 
\scalebox{0.85}{$\begin{bmatrix}
\alpha_{1}e^{-j2\pi n \Delta f\tau_{1}}[\dot{\mathbf{G}}_{1},0]\mathbf{X}[n] \\
+\alpha_{3}e^{-j2\pi n \Delta f\tau_{3}}[\dot{\mathbf{G}}_{3},0]\mathbf{X}[n] \\
+\alpha_{5}e^{-j2\pi n \Delta f\tau_{5}}[0,\dot{\mathbf{G}}_{5}]\mathbf{X}[n] \\
\alpha_{2}e^{-j2\pi n \Delta f\tau_{2}}[\dot{\mathbf{G}}_{2},0]\mathbf{X}[n]
\end{bmatrix}$}, \label{eq:partial_theta1} \\
\frac{\partial \bm{\mu}[n]}{\partial \theta_{2}} &=
\scalebox{0.82}{$\begin{bmatrix}
\alpha_{3}e^{-j2\pi n \Delta f\tau_{3}}[\dot{\mathbf{G}}_{3,\theta_2},0]\mathbf{X}[n]
+\alpha_{5}e^{-j2\pi n \Delta f\tau_{5}}[0,\dot{\mathbf{G}}_{5,\theta_2}]\mathbf{X}[n] \\
\alpha_{2}e^{-j2\pi n \Delta f\tau_{2}}[\dot{\mathbf{G}}_{2,\theta_2},0]\mathbf{X}[n]
+\alpha_{4}e^{-j2\pi n \Delta f\tau_{4}}[0,\dot{\mathbf{G}}_{4,\theta_2}]\mathbf{X}[n]
\end{bmatrix}$}, \label{eq:partial_theta2} \\
\frac{\partial \bm{\mu}[n]}{\partial \alpha_{i}^{R}} &= -j \frac{\partial \bm{\mu}[n]}{\partial \alpha_{i}^{I}} =
\scalebox{0.86}{$\begin{bmatrix}
e^{-j2\pi n \Delta f\tau_{i}}[\mathbf{G}_{i},0]\mathbf{X}[n] \\
\bm{0}
\end{bmatrix}$}. \label{eq:partial_alpha}
\end{align}
Under the resolvable-path/score-orthogonality approximation stated in Section III, the FIM reduces to $\mathbf{J}_{\boldsymbol{\eta}} = \operatorname{blkdiag}[\mathbf{J}_\tau, \mathbf{J}_\theta, \mathbf{J}_\alpha]$. By substituting the partial derivatives \eqref{p.d1}-\eqref{eq:partial_alpha} into the Slepian-Bangs \eqref{S.B}, the diagonal elements of the FIM are obtained as \eqref{eq:J-tau1}-\eqref{eq:J-alpha2}.

The transformation matrix $\mathbf{T}$ is defined as the Jacobian of $\bm{\eta}$ with respect to $\bm{\gamma}$ and can be obtained as:
\begin{equation}
\mathbf{T} \triangleq \frac{\partial \eta}{\partial \gamma} = 
\begin{bmatrix}
\boldsymbol{\Xi} & \mathbf{O} \\
\mathbf{O} & \mathbf{I}_{10}
\end{bmatrix},
\label{eq:T}
\end{equation}
where $\boldsymbol{\Xi} = [{\nabla}_{\bm{\tau},\mathbf{p}}^{T}, {\nabla}_{\bm{\theta},\mathbf{p}}^{T}]^{T}$. Considering the five paths introduced in \eqref{eq:radar_signal} and definition of the parameters $\mathbf{e}_1 = \frac{(\mathbf{p} - \mathbf{q})^T}{\|\mathbf{p} - \mathbf{q}\|}, \quad
\mathbf{e}_2 = \frac{(\mathbf{p} - \mathbf{r})^T}{\|\mathbf{p} - \mathbf{r}\|}$, the gradient expressions with respect to the position $\mathbf{p}$ are given by:
\begin{align}
\nabla_{\tau_1,p} &= \scalebox{0.9}{$\frac{2}{c} \mathbf{e}_{1}$}, \\
\nabla_{\tau_2,p} &= \nabla_{\tau_3,p} = \nabla_{\tau_5,p} = \scalebox{0.9}{$\frac{1}{c} [\mathbf{e}_{1} + \mathbf{e}_{2}]$},\\
\nabla_{\tau_4,p} &= \scalebox{0.9}{$\frac{2}{c} \mathbf{e}_{2}$}.
\end{align}
where parameter $c$ denotes the speed of light in vacuum. The angles $\theta_1$ and $\theta_2$ are computed as: $\theta_1 = \tan^{-1} \left( \frac{y_p - y_q}{x_p - x_q} \right)$ and $\theta_2 = \tan^{-1} \left( \frac{y_p - y_r}{x_p - x_r} \right)$. Thus, the gradient vectors are given by:
\begin{align}
\begin{split}
\nabla_{\theta_{1},\mathbf{p}} &= \scalebox{0.9}{$\left[\frac{y_q - y_p}{(x_p - x_q)^2 + (y_p - y_q)^2},\frac{x_p - x_q}{(x_p - x_q)^2 + (y_p - y_q)^2}\right]$} \\
&= \scalebox{0.9}{$\frac{1}{\|\mathbf{p} - \mathbf{q}\|}\left[-\sin\theta_1,\cos\theta_1\right]$}
=\scalebox{0.9}{$\boldsymbol{\rho}_1^T$},
\end{split}
\label{eq:grad1} \\
\begin{split}
\nabla_{\theta_2,\mathbf{p}}
&= \scalebox{0.9}{$\frac{1}{\|\mathbf{r} - \mathbf{p}\|}\left[-\sin\theta_2,\cos\theta_2\right]$}=
\scalebox{0.9}{$\boldsymbol{\rho}_2^T$}.
\end{split}
\label{eq:grad2}
\end{align}

Therefore, the FIM for $\bm{\gamma}$ is given by:
\begin{equation}
\mathbf{J}_{\bm{\gamma}} = \mathbf{T}^{T}\, \mathbf{J}_{\bm{\eta}} \, \mathbf{T}
=\scalebox{0.9}{$\begin{bmatrix}
\boldsymbol{\Xi}  & \mathbf{O} \\
\mathbf{O} & \mathbf{I}
\end{bmatrix} ^{T}
\mathbf{J}_{\bm{\eta}}
\begin{bmatrix}
\boldsymbol{\Xi}  & \mathbf{O} \\
\mathbf{O} & \mathbf{I}
\end{bmatrix}$}
=\scalebox{0.9}{$\begin{bmatrix}
\boldsymbol{\Xi}^{T} \mathbf{J}_{m} \boldsymbol{\Xi}  & \mathbf{O} \\
\mathbf{O} & \mathbf{J}_{\bm{\alpha}}
\end{bmatrix}$},
\end{equation}
where $\mathbf{J}_{\boldsymbol{m}} = \operatorname{blkdiag}[\mathbf{J}_\tau, \mathbf{J}_\theta]$. 
We can also obtain:
\begin{align}
\boldsymbol{\Xi} &= \scalebox{0.9}{$\left[ 2c^{-1}\mathbf{e}_1, c^{-1}\mathbf{e}_1 + c^{-1}\mathbf{e}_2, c^{-1}\mathbf{e}_1 + c^{-1}\mathbf{e}_2, \right.$} \nonumber \\
&\quad \scalebox{0.9}{$\left. 2c^{-1}\mathbf{e}_2, c^{-1}\mathbf{e}_1 + c^{-1}\mathbf{e}_2, \boldsymbol{\rho}_1, \boldsymbol{\rho}_2 \right]^{T}$}.
\end{align}

Therefore, the position-related submatrix denoted as effective FIM is derived as follows:
\begin{align}
\mathbf{J}_{e_{2\times 2}} &= \bm{\Xi}^{T}\mathbf{J}_{m}\bm{\Xi} \nonumber \\
&= \scalebox{0.9}{$J_{\tau_{1}\tau_{1}}\frac{4}{c^{2}}\mathbf{e}_{1}\mathbf{e}_{1}^{T}+J_{\tau_{2}\tau_{2}}\frac{1}{c^{2}}[\mathbf{e}_{1}+\mathbf{e}_{2}][\mathbf{e}_{1}+\mathbf{e}_{2}]^{T}$} \nonumber \\
&\quad \scalebox{0.9}{$+ J_{\tau_{3}\tau_{3}}\frac{1}{c^{2}}[\mathbf{e}_{1}+\mathbf{e}_{2}][\mathbf{e}_{1}+\mathbf{e}_{2}]^{T}+J_{\tau_{4}\tau_{4}}\frac{4}{c^{2}}\mathbf{e}_{2}\mathbf{e}_{2}^{T}$} \nonumber \\
&\quad \scalebox{0.9}{$+ J_{\tau_{5}\tau_{5}}\frac{1}{c^{2}}[\mathbf{e}_{1}+\mathbf{e}_{2}][\mathbf{e}_{1}+\mathbf{e}_{2}]^{T}$} \nonumber \\
&\quad \scalebox{0.9}{$+ J_{\theta_{1}\theta_{1}}\boldsymbol{\rho}_{1}\boldsymbol{\rho}_{1}^{T}+J_{\theta_{2}\theta_{2}}\boldsymbol{\rho}_{2}\boldsymbol{\rho}_{2}^{T}$}.
\end{align}

Substituting the individual FIM elements, we can obtain:
\begin{subequations}
\begin{align}
\mathbf{J}_{e} &= \scalebox{0.9}{$\frac{2N_s(N_s^{2}-1)}{12\sigma^{2}c^{2}}(2\pi\Delta f)^{2}$} \times \nonumber \\
&\quad \scalebox{0.9}{$\Bigg[ |\alpha_{1}|^{2}\mathrm{tr}(\mathbf{G}_{1}\mathbf{R}_{X1}\mathbf{G}_{1}^{H})4\mathbf{e}_{1}\mathbf{e}_{1}^{T}$} \nonumber \\
&\quad \scalebox{0.9}{$+ |\alpha_{2}|^{2}\mathrm{tr}(\mathbf{G}_{2}\mathbf{R}_{X1}\mathbf{G}_{2}^{H})(\mathbf{e}_{1}+\mathbf{e}_{2})(\mathbf{e}_{1}+\mathbf{e}_{2})^{T}$} \nonumber \\
&\quad \scalebox{0.9}{$+ |\alpha_{3}|^{2}\mathrm{tr}(\mathbf{G}_{3}\mathbf{R}_{X1}\mathbf{G}_{3}^{H})(\mathbf{e}_{1}+\mathbf{e}_{2})(\mathbf{e}_{1}+\mathbf{e}_{2})^{T}$} \nonumber \\
&\quad \scalebox{0.9}{$+ 4|\alpha_{4}|^{2}\mathrm{tr}(\mathbf{G}_{4}\mathbf{R}_{X2}\mathbf{G}_{4}^{H})\mathbf{e}_{2}\mathbf{e}_{2}^{T}$} \nonumber \\
&\quad \scalebox{0.9}{$+ |\alpha_{5}|^{2}\mathrm{tr}(\mathbf{G}_{5}\mathbf{R}_{X2}\mathbf{G}_{5}^{H})(\mathbf{e}_{1}+\mathbf{e}_{2})(\mathbf{e}_{1}+\mathbf{e}_{2})^{T} \Bigg]$} \nonumber \\
&\quad + \scalebox{0.9}{$\Bigg[ \frac{2}{\sigma^{2}}|\alpha_{1}|^{2}\mathrm{tr}\{\dot{\mathbf{G}}_{1,\theta_{1}}\mathbf{R}_{X1}\dot{\mathbf{G}}_{1,\theta_{1}}^{H}\}N_s$} \nonumber \\
&\quad \scalebox{0.9}{$+ \frac{2}{\sigma^{2}}|\alpha_{3}|^{2}\mathrm{tr}\{\dot{\mathbf{G}}_{3,\theta_{1}}\mathbf{R}_{X1}\dot{\mathbf{G}}_{3,\theta_{1}}^{H}\}N_s$} \nonumber \\
&\quad \scalebox{0.9}{$+ \frac{2}{\sigma^{2}}|\alpha_{5}|^{2}\mathrm{tr}\{\dot{\mathbf{G}}_{5,\theta_{1}}\mathbf{R}_{X2}\dot{\mathbf{G}}_{5,\theta_{1}}^{H}\}N_s$} \nonumber 
\end{align}
\begin{align}
&\quad \scalebox{0.9}{$+ \frac{2}{\sigma^{2}}|\alpha_{2}|^{2}\mathrm{tr}\{\dot{\mathbf{G}}_{2,\theta_{1}}\mathbf{R}_{X1}\dot{\mathbf{G}}_{2,\theta_{1}}^{H}\}N_s\Bigg]\boldsymbol{\rho}_{1}\boldsymbol{\rho}_{1}^{T}$} \nonumber \\
&\quad + \scalebox{0.9}{$\Bigg[ \frac{2}{\sigma^{2}}|\alpha_{5}|^{2}\mathrm{tr}\{\dot{\mathbf{G}}_{5,\theta_{2}}\mathbf{R}_{X2}\dot{\mathbf{G}}_{5,\theta_{2}}^{H}\}N_s$} \nonumber \\
&\quad \scalebox{0.9}{$+ \frac{2}{\sigma^{2}}|\alpha_{2}|^{2}\mathrm{tr}\{\dot{\mathbf{G}}_{2,\theta_{2}}\mathbf{R}_{X1}\dot{\mathbf{G}}_{2,\theta_{2}}^{H}\}N_s$} \nonumber \\
&\quad \scalebox{0.9}{$+ \frac{2}{\sigma^{2}}|\alpha_{3}|^{2}\mathrm{tr}\{\dot{\mathbf{G}}_{3,\theta_{2}}\mathbf{R}_{X1}\dot{\mathbf{G}}_{3,\theta_{2}}^{H}\}N_s$} \nonumber \\
&\quad \scalebox{0.9}{$+ \frac{2}{\sigma^{2}}|\alpha_{4}|^{2}\mathrm{tr}\{\dot{\mathbf{G}}_{4,\theta_{2}}\mathbf{R}_{X2}\dot{\mathbf{G}}_{4,\theta_{2}}^{H}\}N_s\Bigg]\boldsymbol{\rho}_{2}\boldsymbol{\rho}_{2}^{T}$}. \label{eq:Je_final}
\end{align}
\end{subequations}

Then, we define new parameters as follows:
\begin{align}
\lambda_{k} &\triangleq \scalebox{0.9}{$\frac{2\pi^{2}N_s(N_s^{2}-1)(\Delta f)^{2}}{3\sigma^{2}c^{2}}|\alpha_{k}|^{2}l_{k}$}, \qquad \scalebox{0.9}{$k=1,2,\ldots,5$} \label{eq:lambda}
\end{align}
where $l_1 = 4\operatorname{tr}\{\mathbf{G}_1 \mathbf{R}_{x1} \mathbf{G}_1^{H}\}$, $l_2 = \operatorname{tr}\{\mathbf{G}_2 \mathbf{R}_{x1} \mathbf{G}_2^{H}\}$, $l_3 = \operatorname{tr}\{\mathbf{G}_3 \mathbf{R}_{x1} \mathbf{G}_3^{H}\}$, $l_4 = 4\operatorname{tr}\{\mathbf{G}_4 \mathbf{R}_{x2} \mathbf{G}_4^{H}\}$ and $l_5 = \operatorname{tr}\{\mathbf{G}_5 \mathbf{R}_{x2} \mathbf{G}_5^{H}\}$.

\begin{align}
\delta_{k} &\triangleq \scalebox{0.9}{$\frac{2}{\sigma^{2}}N_s|\alpha_{k}|^{2}m_{k}$}, \quad \scalebox{0.9}{$k=1,2,3,5$}. 
\end{align}
where $m_k = \operatorname{tr}\{\dot{\mathbf{G}}_{k,\theta_1} \mathbf{R}_x^{(k)} \dot{\mathbf{G}}_{k,\theta_1}^{H}\}$, with $\mathbf{R}_x^{(k)} = \mathbf{R}_{x1}$ for $k \in \{1,2,3\}$ and $\mathbf{R}_x^{(k)} = \mathbf{R}_{x2}$ for $k = 5$.
\begin{align}
\eta_{k}  &\triangleq \scalebox{0.9}{$\frac{2}{\sigma^{2}}N_s|\alpha_{k}|^{2}n_{k}$}, \quad \scalebox{0.9}{$k=2,3,4,5$} 
\end{align}
where
$n_{2} = \mathrm{tr}\{\dot{\mathbf{G}}_{2,\theta_{2}}\mathbf{R}_{x1}\dot{\mathbf{G}}_{2,\theta_{2}}^{H}\},$
$n_{3} = \mathrm{tr}\{\dot{\mathbf{G}}_{3,\theta_{2}}\mathbf{R}_{x1}\dot{\mathbf{G}}_{3,\theta_{2}}^{H}\},$
$n_{4} = \mathrm{tr}\{\dot{\mathbf{G}}_{4,\theta_{2}}\mathbf{R}_{x2}\dot{\mathbf{G}}_{4,\theta_{2}}^{H}\}$ and
$n_{5} = \mathrm{tr}\{\dot{\mathbf{G}}_{5,\theta_{2}}\mathbf{R}_{x2}\dot{\mathbf{G}}_{5,\theta_{2}}^{H}\}$. Thus, the equivalent FIM is:
\begin{align}
\mathbf{J}_{e} &= \scalebox{0.9}{$(\lambda_{1}+\lambda_{2}+\lambda_{3}+\lambda_{5})\mathbf{e}_{1}\mathbf{e}_{1}^{T}$} \nonumber \\
&\quad \scalebox{0.9}{$+(\lambda_{2}+\lambda_{3}+\lambda_{4}+\lambda_{5})\,\mathbf{e}_{2}\mathbf{e}_{2}^{T}$} \nonumber \\
&\quad \scalebox{0.9}{$+(\lambda_{2}+\lambda_{3}+\lambda_{5})\,\mathbf{e}_{1}\mathbf{e}_{2}^{T}+(\lambda_{2}+\lambda_{3}+\lambda_{5})\,\mathbf{e}_{2}\mathbf{e}_{1}^{T}$} \nonumber \\
&\quad \scalebox{0.9}{$+(\delta_{1}+\delta_{2}+\delta_{3}+\delta_{5})\,\boldsymbol{\rho}_{1}\boldsymbol{\rho}_{1}^{T}+(\eta_{2}+\eta_{3}+\eta_{4}+\eta_{5})\,\boldsymbol{\rho}_{2}\boldsymbol{\rho}_{2}^{T}$}.
\end{align}

Then, we use additional scalar factor as follows: $L_{i} \triangleq \scalebox{0.9}{$\frac{2\pi^{2}N_s(N_s^{2}-1)}{3\sigma^{2}c^{2}}(\Delta f)^{2}|\alpha_{i}|^{2}$}$ and 
$\mathbf{M}_{i} \triangleq \mathbf{G}_{i}^{H}\mathbf{G}_{i}$.
Thus, we can obtain:
\begin{align}
\lambda_{1} &= \scalebox{0.9}{$\frac{8\pi^{2}N_s(N_s^{2}-1)}{3\sigma^{2}c^{2}}(\Delta f)^{2}|\alpha_{1}|^{2}\mathrm{tr}\{\mathbf{G}_{1}\mathbf{R}_{x1}\mathbf{G}_{1}^{H}\}$} \nonumber \\
&= \scalebox{0.9}{$4L_{1}\,\mathrm{tr}\{\mathbf{M}_{1}\mathbf{R}_{x1}\}$},
\end{align}
Analogous expressions can be derived $\lambda_{2} = L_{2}\mathrm{tr}\{\mathbf{M}_{2}\mathbf{R}_{X1}\},$
$\lambda_{3} = L_{3}\mathrm{tr}\{\mathbf{M}_{3}\mathbf{R}_{X1}\},$
$\lambda_{4} = 4L_{4}\mathrm{tr}\{\mathbf{M}_{4}\mathbf{R}_{X2}\}$ and
$\lambda_{5} = L_{5}\mathrm{tr}\{\mathbf{M}_{5}\mathbf{R}_{X2}\}$. Thus the equivalent FIM is:
\begin{flalign}
\mathbf{J}_e &= \begin{bmatrix}
J_{1,1} & J_{1,2} \\
J_{1,2} & J_{2,2}
\end{bmatrix} = \boldsymbol{\Xi}^{T} \mathbf{J}_{m} \boldsymbol{\Xi}\nonumber \\
&= \scalebox{0.9}{$\big( 4 L_1\operatorname{tr}\{\mathbf{M}_1\mathbf{R}_{x1}\}  
+ L_2\operatorname{tr}\{\mathbf{M}_2\mathbf{R}_{x1}\}$} + L_3\operatorname{tr}\{\mathbf{M}_3\mathbf{R}_{x1}\} 
\nonumber \\
&\quad \scalebox{0.9}{$+ 4L_4\operatorname{tr}\{\mathbf{M}_4\mathbf{R}_{x2}\}+ L_5\operatorname{tr}\{\mathbf{M}_5\mathbf{R}_{x2}\}\big)
\mathbf{e}_{1}\mathbf{e}_{1}^{T}$}
\nonumber \\[4pt]
&\quad + \scalebox{0.9}{$\big( L_2\operatorname{tr}\{\mathbf{M}_2\mathbf{R}_{x1}\}
+ L_3\operatorname{tr}\{\mathbf{M}_3\mathbf{R}_{x1}\}$} \nonumber \\
&\quad \scalebox{0.9}{$+ 4L_4\operatorname{tr}\{\mathbf{M}_4\mathbf{R}_{x2}\}
+ L_5\operatorname{tr}\{\mathbf{M}_5\mathbf{R}_{x2}\}\big)
\mathbf{e}_{2}\mathbf{e}_{2}^{T}$}
+ \hdots
\end{flalign}
After collecting the coefficients of the feasible covariance vector $\mathbf r=[\mathbf r_1^T,\mathbf r_2^T]^T$, each real EFIM entry is linear in $\mathbf r$. Hence, there exist coefficient vectors $\mathbf c_{11}$, $\mathbf c_{22}$, and $\mathbf c_{12}$ such that
\begin{equation}
J_{11}=\mathbf c_{11}^H\mathbf r,\qquad J_{22}=\mathbf c_{22}^H\mathbf r,\qquad J_{12}=\mathbf c_{12}^H\mathbf r,
\end{equation}
with real values on the feasible Hermitian covariance set. The coefficient vectors contain the contributions of both $\mathbf r_1$ and $\mathbf r_2$. Defining
\begin{equation}
\mathbf m\triangleq\frac{\mathbf c_{11}+\mathbf c_{22}}{2},\qquad
\widetilde{\mathbf B}\triangleq\frac{\mathbf c_{11}\mathbf c_{22}^H+\mathbf c_{22}\mathbf c_{11}^H}{2}-\mathbf c_{12}\mathbf c_{12}^H,
\label{eq:ND}
\end{equation}
we obtain
\begin{equation}
\operatorname{CRB}=\frac{J_{11}+J_{22}}{J_{11}J_{22}-J_{12}^2}
=\frac{\mathbf r^H\mathbf m+\mathbf m^H\mathbf r}{\mathbf r^H\widetilde{\mathbf B}\mathbf r}.
\label{eq:CRB-end}
\end{equation}
The off-diagonal blocks of $\widetilde{\mathbf B}$ explicitly generate the $\mathbf r_1$--$\mathbf r_2$ cross terms. No positive-semidefiniteness of $\widetilde{\mathbf B}$ is assumed, which motivates the spectral split used in Section IV-B.

\section{Calculation of the CRB in terms of $\phi$}
\label{app:B}

For fixed $\mathbf A$ and $\mathbf A_a$, the phase vector affects the radar FIM through Path~3. Define
\begin{align}
\mathbf d_0&\triangleq\overline{\mathbf c(\theta_4)}\odot(\mathbf I-\mathbf A)\mathbf c(\theta_5),\\
\mathbf d_1&\triangleq\overline{\mathbf c(\theta_4)}\odot(\mathbf I-\mathbf A)\dot{\mathbf c}(\theta_5),
\end{align}
where $\dot{\mathbf c}(\theta_5)=\partial\mathbf c(\theta_5)/\partial\theta_2$ and $\partial\theta_5/\partial\theta_2=1$. From \eqref{eq:gain_g},
\begin{equation}
g=\mathbf d_0^T\boldsymbol\phi,\qquad
\frac{\partial g}{\partial\theta_2}=\mathbf d_1^T\boldsymbol\phi.
\end{equation}
Hence
\begin{equation}
S_0(\boldsymbol\phi)=|g|^2=\boldsymbol\phi^H\mathbf Q_0\boldsymbol\phi,\qquad
S_1(\boldsymbol\phi)=\left|\frac{\partial g}{\partial\theta_2}\right|^2=\boldsymbol\phi^H\mathbf Q_1\boldsymbol\phi,
\end{equation}
where $\mathbf Q_0=\mathbf d_0^*\mathbf d_0^T$ and $\mathbf Q_1=\mathbf d_1^*\mathbf d_1^T$. The Path-3 delay information and its $\theta_1$ contribution are affine in $S_0$, whereas the Path-3 contribution to the $\theta_2$ information is affine in $S_1$. Consequently, the position-EFIM entries can be written as
\begin{equation}
J_{ij}(\boldsymbol\phi)=\beta^{(0)}_{ij}S_0(\boldsymbol\phi)+\beta^{(1)}_{ij}S_1(\boldsymbol\phi)+\lambda_{ij},
\end{equation}
for $(i,j)\in\{(1,1),(1,2),(2,2)\}$, where the coefficients collect the phase-independent terms. Therefore,
\begin{equation}
\operatorname{CRB}(\boldsymbol\phi)=F\!\left(S_0(\boldsymbol\phi),S_1(\boldsymbol\phi)\right)
=\frac{J_{11}(\boldsymbol\phi)+J_{22}(\boldsymbol\phi)}{J_{11}(\boldsymbol\phi)J_{22}(\boldsymbol\phi)-J_{12}^2(\boldsymbol\phi)}.
\label{CRB-phi}
\end{equation}
At iteration $t$, the local SCA model used in Section IV-C is obtained from the first-order expansion
\begin{equation}
F(S_0,S_1)\approx F(s_{0,t},s_{1,t})+\mu_{0,t}(S_0-s_{0,t})+\mu_{1,t}(S_1-s_{1,t}),
\end{equation}
where $s_{q,t}=S_q(\boldsymbol\phi^{(t)})$ and $\mu_{q,t}=\partial F/\partial S_q|_{(s_{0,t},s_{1,t})}$, $q\in\{0,1\}$. The acceptance rule in Section IV-C is applied to the original fixed-penalty objective rather than treating this local expansion as a global MM upper bound.

%\vfill\pagebreak
\bibliographystyle{IEEEtran}
\balance
\bibliography{refs}

\end{document}